\documentclass[floatfix,twocolumn,showpacs,preprintnumbers,amsmath,amssymb,pra,superscriptaddress,longbibliography]{revtex4-1}
\usepackage{color}
\usepackage[usenames,dvipsnames,svgnames,table]{xcolor}
\usepackage[colorlinks=true,linkcolor=blue,urlcolor=blue,citecolor=blue]{hyperref}
\usepackage{mathtools}
\usepackage{graphicx}
\usepackage{dcolumn}
\usepackage{array}
\usepackage{lipsum}
\usepackage{bm}
\usepackage{subfigure}
\usepackage{amssymb}
\usepackage{multirow}
\usepackage{tabularx}
\usepackage{amsmath}
\usepackage{braket}
\usepackage{csquotes}
\graphicspath{{plots/}}
 \usepackage{lipsum}
\usepackage{mathrsfs}
\usepackage{MnSymbol}
	
\newcommand{\beq}{\begin{equation}}
\newcommand{\eeq}{\end{equation}}
\newcommand{\bea}{\begin{eqnarray}}
\newcommand{\eea}{\end{eqnarray}}

\begin{document}

\title{Dielectric formalism of the 2D uniform electron gas at finite temperatures}
\author{Fotios Kalkavouras}
 \affiliation{Department of Electromagnetics and Plasma Physics, Royal Institute of Technology (KTH), Stockholm, SE-100 44, Sweden}
\author{Tobias Dornheim}
 \affiliation{Institute of Radiation Physics, Helmholtz-Zentrum Dresden-Rossendorf (HZDR), D-01328 Dresden, Germany}
 \affiliation{Center for Advanced Systems Understanding (CASUS) at Helmholtz-Zentrum Dresden-Rossendorf (HZDR), D-02826 G\"orlitz, Germany}
\author{Paul Hamann}
 \affiliation{Institute of Radiation Physics, Helmholtz-Zentrum Dresden-Rossendorf (HZDR), D-01328 Dresden, Germany}
 \affiliation{Institut f\"ur Physik, Universit\"at Rostock, D-18057 Rostock, Germany}
\author{Panagiotis Tolias}\email{tolias@kth.se}
 \affiliation{Department of Electromagnetics and Plasma Physics, Royal Institute of Technology (KTH), Stockholm, SE-100 44, Sweden}

\begin{abstract}
\noindent We present a comprehensive analysis of the two-dimensional uniform electron gas (2D-UEG or more commonly 2DEG) at finite temperature, spanning a broad range of densities / coupling strengths ($0.01\le{r}_s\le20$) and temperatures / degeneracy parameters ($0.01\le\Theta= k_\textnormal{B}T/E_\textnormal{F} \le 10$). Within the self-consistent dielectric formalism, we construct two-dimensional versions of the Singwi--Tosi--Land--Sj\"olander (STLS)  and hypernetted-chain (HNC) approximation based schemes. We benchmark the accuracy of the STLS and the HNC schemes against new state-of-the-art path-integral Monte Carlo data. We also report structural and thermodynamic properties across the full $(r_s,\Theta)$ phase diagram domain studied, identify regimes in which these schemes remain quantitatively reliable, and provide an accurate parametrization of the exchange--correlation free energy of the finite-temperature 2DEG.
\end{abstract}
\maketitle

\section{Introduction}

\noindent The uniform electron gas (UEG) (also referred to as quantum one-component plasma, homogeneous electron gas, uniform electron liquid or jellium) constitutes one of the central model systems of many-body physics~\cite{quantum_theory}. Being the archetypal system of fermions correlated through long-range Coulomb interactions, it has played a foundational role in the development of electronic structure theory, many-body perturbation theory, and density functional theory (DFT)~\cite{quantum_theory,Jones_RMP_2015,loos,mahan1990many}. Despite its apparent simplicity, the UEG exhibits a rich interplay between classical and quantum effects, making it an ideal model system for both computational and analytical approaches~\cite{Ichimaru_1993,review,loos,Shepherd_JCP_2012,Schoof_CPP_2015,Malone_JCP_2015,dornheim_dynamic,Takada_PRB_2016}.

In \emph{three dimensions}, the equilibrium UEG properties have been investigated extensively over several decades. Ground state quantum Monte Carlo (QMC) simulations have provided highly accurate benchmark data for its thermodynamic and structural properties~\cite{Ceperley_Alder_PRL_1980,Spink_PRB_2013,Ortiz_PRB_1994,Ortiz_PRL_1999}, which in turn underpin exchange--correlation parametrizations that are widely utilized in DFT~\cite{Perdew_Wang,Perdew_Zunger_PRB_1981,vwn}. Interest soon expanded to finite temperatures~\cite{Khanna_1976,Gouedard_1978,Gupta_1980,Gupta_1982,Ichimaru_1982,Arista_1984,Perrot_1984,Pokrant_1977,Dandrea_1986,Ichimaru_1987} with a more recent focus on the warm dense matter (WDM) regime~\cite{review,new_POP,Dornheim_review,vorberger2025roadmapwarmdensematter}. In this context, large-scale path integral Monte Carlo (PIMC) simulations~\cite{Brown_PRL_2013,Schoof_PRL_2015,dornheim_prl,groth_prl,dornheim_ML} have enabled the systematic quasi-exact analysis of the thermodynamic, structural, and dynamic properties of the finite-temperature 3D-UEG. Furthermore, different flavors of the self-consistent dielectric formalism~\cite{Nozieres_1958,SingwiTosi_Review,Ichimaru_1987} such as the Singwi--Tosi--Land--Sj\"olander (STLS) scheme~\cite{stls_original,stls}, the hypernetted-chain (HNC) approximation based scheme~\cite{tanaka_hnc,dornheim_electron_liquid} and their various extensions~\cite{vs_original,qstls_original,Tolias_JCP_2021,Tolias_JCP_2023}, have proven invaluable for describing the static and even dynamic linear response properties~\cite{Tolias_JCP_2024,Tolias_PRB_2024,Tolias_2025}. Nowadays, dielectric schemes are also utilized in close synergy with PIMC simulations~\cite{review,Dornheim_review}. In particular, modern exchange--correlation parametrizations incorporate finite-temperature information and, in some cases, explicitly rely on dielectric theory input to capture temperature-dependent correlation effects~\cite{groth_prl,review}.

In \emph{two dimensions}, in stark contrast to the 3D case, the UEG literature remains rather sparse. To our knowledge, accurate QMC data have been thus far restricted to the ground state limit~\cite{Tanatar_1989,Drummond_2009,Drummond_2DEG_2013,Smith_2DEG_2024}, while finite temperature effects have been scantily analyzed with the dielectric formalism~\cite{Ando_1982,Maldague_1978,Schweng_1994,Yurtsever_2003,Khanh_2007,Bhukal_2019} employing the STLS scheme as well as the quantum-classical mapping approach~\cite{Perrot_2DEG_2001,Dharmawardana_2DEG_2003,Khanh_2004,Dharmawardana_2DEG_2012} employing the original Perrot \& Dharma-wardana formulation~\cite{perrot} that features empirical elements. Therefore, quasi-exact PIMC reference data are essentially absent for the finite temperature 2DEG and the applicability of dielectric schemes to low dimensional Coulomb systems remains poorly tested. This gap is particularly noteworthy given the growing relevance of thermal excitations in 2D electronic systems. Ultrafast laser excitation~\cite{Brida_2013,Jensen_2014,Tielrooij_2013,Sun_2008}, photo-induced transport~\cite{Strait_2008,Shabani_2016}, and electronic heating in high-mobility semiconductor heterostructures~\cite{Ando_1982} and atomically thin materials routinely drive electronic temperatures far above the lattice temperature~\cite{Sarma_2004}. This applies to a wide range of platforms, including quantum wells, transition-metal dichalcogenides~\cite{Wang_2012}, graphene, electrons on helium~\cite{Yuriy_2007}, many of which allow a continuous tuning between the weakly and strongly correlated regimes.

In this work, we address this deficiency by extending different schemes of the self-consistent dielectric formalism to the finite temperature 2DEG and by benchmarking them against new state-of-the-art PIMC results. In particular, we focus on the STLS scheme, which is the prototypical weak-to-moderate coupling dielectric theory that is perturbative in nature and based on truncating the BBGKY hierarchy, and on the HNC scheme, which is the prototypical moderate-to-strong coupling dielectric theory that is non-perturbative in nature and based on the integral equation theory of liquids. For the STLS and HNC dielectric schemes, we compute structural and thermodynamic properties across a broad range of density \& temperature and we identify the regimes in which they remain quantitatively reliable. For the STLS scheme, we also provide accurate parametrizations of the interaction energy and exchange--correlation free energy.

The paper is organized as follows. In Secs.~\ref{sec:theory}, \ref{sec:Dielectric}, we introduce basic theoretical quantities for the 2DEG and outline the general computational procedure that characterizes all dielectric schemes regardless of dimensionality and of complexity. In Secs.~\ref{subsub1} and \ref{subsub2}, the 2D versions of the STLS and HNC schemes are presented along with their computationally convenient normalized system of equations. In Sec.~\ref{sec:PIMC}, the general PIMC simulation procedure is outlined. In Secs.~\ref{sec:SSF}, \ref{sec:pcf} and \ref{sec:chi}, we introduce results for the static structure factor, the pair correlation function and the static density response function as well as compare dielectric scheme predictions to the PIMC data. Sec.~\ref{sec:uintfxc} presents the interaction energies and exchange--correlation free energies alongside their accurate finite-temperature parametrizations. The paper is concluded by a summary and outlook in Sec.~\ref{sec:outlook}.

\section{Theory\label{sec:theory}}

\noindent The 2DEG, just as its 3D counterpart, is a spatially homogeneous quantum model system of electrons embedded in a rigid uniform charge neutralizing background~\cite{quantum_theory} that is characterized by the uniform areal charge density of $-ne$, where $n$ is the electron density. In the standard two dimensional analogue, where electrons are restricted to a plane, the Coulomb interaction retains its $\propto1/r$ form but it acquires a $\propto1/k$ reciprocal space form~\cite{Baus_Hansen_OCP,LuccoCastello_2DEG_2021}. There is an alternative two dimensional analogue, where the pair potential is derived from the solution of the 2D Poisson's equation for a point charge source, which leads to the standard $\propto1/k^2$ reciprocal space form, but to a logarithmic $\propto\ln(r)$ real space form~\cite{Jancovici_1981,Caillol_1982,Khrapak_2DEGlog_2016}. Throughout this work, we naturally consider the first analogue. 

The thermodynamic state of the 2DEG is fully specified by three dimensionless parameters~\cite{quantum_theory,review,Schweng_1994}: (1) The \emph{quantum coupling parameter} $r_s=d/a_\textnormal{B}$, with $d=1/\sqrt{\pi{n}}$ the 2D Wigner--Seitz radius and with $a_\textnormal{B}=\hbar^2/(m_e e^2)$ the first Bohr radius. (2) The \emph{quantum degeneracy parameter} $\Theta=k_\textnormal{B}T/{E_\textnormal{F}}$, with $E_\textnormal{F}=(\hbar^2q_\textnormal{F}^2)/(2 m_e)$ the Fermi energy $E_\textnormal{F}$ and $q_\textnormal{F} = \sqrt{2\pi n^{\uparrow}}$ the 2D spin-up Fermi wavevector. (3) The \emph{spin polarization parameter} $\zeta=(n^\uparrow - n^\downarrow)/n$, for which $0\leq\zeta\leq1$. In this work, we restrict ourselves to the paramagnetic (i.e., unpolarized) case of $\zeta=0$.

In the high density limit $r_s \to 0$, the 2DEG approaches the non-interacting (i.e., ideal) 2D Fermi gas. The general $D$-dimensional expression for the corresponding ideal (Lindhard) density response function reads as~\cite{quantum_theory}
\begin{equation*}
\chi_{0,D}(\mathbf{q},\omega)
= -\frac{2}{V} \sum_{\mathbf{k} }^{(D)}
\frac{ f_0(\mathbf{k}+\mathbf{q}) - f_0(\mathbf{k}) }
{\hbar\omega - \epsilon(\mathbf{k}+\mathbf{q}) + \epsilon(\mathbf{k}) + \imath0} ,
\end{equation*}
where $V$ denotes the $D$-dimensional system volume, with
\begin{equation*}
    \epsilon(q) = \frac{\hbar^2 q^2}{2 m_e}, \qquad  f_0(q) = \frac{1}{\exp\!\left( \frac{\hbar^2 q^2}{2 m_e k_BT} - \bar{\mu} \right) + 1 } ,
\end{equation*}
the electron kinetic energy and Fermi-Dirac distribution, respectively. Additionally, $\sum^{(D)}_{\mathbf{k}}$ denotes a sum over $D$-dimensional wave vectors. The general replacement rule $V^{-1}\sum_{\mathbf{k}}^{(D)}\to\int d^Dk(2\pi)^{-D}$ allows one to switch from the finite periodic particle system to the $D$-dimensional thermodynamic limit leading to
\begin{equation*}
    \chi_{0,D}(\mathbf{q},\omega)
    = -2 \int \frac{d^D k}{(2\pi)^D}
    \frac{  f_0(\mathbf{k}+\mathbf{q}) -f_0(\mathbf{k})}
    { \hbar\omega - \epsilon(\mathbf{k}+\mathbf{q}) + \epsilon(\mathbf{k})  + \imath0},
\end{equation*}
Finally, the dimensionless chemical potential $\bar{\mu}=\beta\mu$ is fixed by the normalization condition
\begin{equation*}
    \int \frac{d^D k}{(2\pi)^D} f_0(\mathbf{k}) = \frac{n}{2}.
\end{equation*}

\subsection{Dielectric formalism \label{sec:Dielectric}}

\noindent The self-consistent dielectric formalism is an established highly accurate microscopic framework for the description of the thermodynamic and structural properties of interacting many-body systems~\cite{Nozieres_1958,SingwiTosi_Review}. In 3D, multiple dielectric schemes have been extensively benchmarked against first-principles UEG data and have been shown to capture correlation, exchange, and finite-temperature effects with remarkable fidelity~\cite{Nozieres_1958,stls_original,Ichimaru_1993,review,dornheim_electron_liquid,Tolias_JCP_2021,Tolias_JCP_2023,Dornheim_review}. Its construction, which combines linear response theory with approximate closures motivated either by kinetic theory~\cite{bonitz_book} or integral equation theory and memory function approaches~\cite{hansen2013theory}, is not restricted to a particular dimension. Thus, extending the dielectric formalism to two dimensions is a reasonable progression, enabling a systematic and internally consistent treatment of exchange-correlation effects in the finite-temperature 2DEG.

All finite temperature dielectric formalism schemes, regardless of complexity and dimensionality, at their base are composed of \textit{three elementary steps}. First, the Matsubara series for the static structure factor (SSF), stemming from the quantum fluctuation-dissipation theorem,  
\begin{equation}
    S(\mathbf{q}) = -\frac{1}{n\beta} \sum_{\ell=-\infty}^{+\infty} \widetilde{\chi}_D \left( \mathbf{q},  i \omega_{\ell} \right).
\label{eq:SSF}
\end{equation}
where $\widetilde{\chi}_D(\mathbf{q},i\omega_\ell)$ is the density response function analytically continued to the complex frequency plane with $\omega_\ell=2\pi\ell/(\beta\hbar)$ the imaginary bosonic Matsubara frequencies~\cite{stls}. Second, the polarization approach expression for the Matsubara density response function $\widetilde{\chi}_D(\mathbf{q},i\omega_\ell)$ that introduces local field corrections to the random phase approximation (RPA)~\cite{Ichimaru_1982}
\begin{equation}
\widetilde{\chi}_D(\mathbf{q},i\omega_\ell)=\frac{\widetilde{\chi}_{0,D}(\mathbf{q},i\omega_\ell)}{1-U_D(\mathbf{q})\left[1-G_D(\mathbf{q},i\omega_\ell)\right]\widetilde{\chi}_{0,D}(\mathbf{q},i\omega_\ell)}\,,
\label{eq:DRF}
\end{equation}
where $U_D(\mathbf{q})$ is the Fourier transformed Coulomb interaction, $U_{2D}(\mathbf{q})=2\pi e^2/q$, and $G_D(\mathbf{q},i\omega_\ell)$ is the dynamic local field correction (LFC) in Matsubara space, which incorporates Pauli exchange, quantum diffraction, and correlation effects beyond the mean field description~\cite{Kugler1975}. Evidently, setting $G_D(\mathbf{q},i\omega_\ell)=0$ recovers the RPA, i.e., mean field. Third, the functional relation between the LFC and SSF
\begin{equation}
    G_D(\mathbf{q},i\omega_\ell)=F_{\mathrm{G}}[S(\mathbf{q})].
    \label{eq:LFC-SSF}
\end{equation}
Combining Eqs.(\ref{eq:SSF},\ref{eq:DRF},\ref{eq:LFC-SSF}), we obtain the functional relation
\begin{equation*}
    S(\mathbf{q}) = -\frac{1}{n \beta} \sum_{\ell=-\infty}^{+\infty} \frac{\widetilde{\chi}_{0,D}\left(\mathbf{q}, i\omega_\ell\right)}{1 - U_D(\mathbf{q}) \{1 - F_{\mathrm{G}}[S(\mathbf{q})]\} \widetilde{\chi}_{0,D}\left(\mathbf{q}, i\omega_\ell\right)}.
    \label{eq:SSF2}
\end{equation*}
This allows the determination of the SSF in an iterative fashion. 

The only differentiating factor between multiple dielectric schemes is the LFC functional $F_{\mathrm{G}}[S]$. In fact, a wide range of dielectric schemes have been developed over the past decades, differing in how the a-priori unknown LFC is being approximated. Most schemes fall into two broad categories. The first category concerns early truncations of the classical or quantum BBGKY hierarchy, which lead to semiclassical STLS-like schemes (static LFC) or quantum STLS-like extensions (dynamic LFC) where the LFC is generated self-consistently\,\cite{stls,stls_original,vs_original,Tolias_PRB_2024,qstls_original,Holas_PRB_1987,Tolias_2025}. The second category concerns classical manipulations based on the integral equation theory of liquids followed by ad hoc inclusion of quantum effects based on a Vlasov-to-Lindhard correspondence rule and often also a STLS-to-qSTLS correspondence rule, which provides a less rigorous but yet non-perturbative route to constructing the LFC from correlations encoded in the Ornstein--Zernike (OZ) equation and its approximate non-linear closures (e.g. HNC)\,\cite{tanaka_hnc,dornheim_electron_liquid,Tolias_JCP_2021,LuccoCastello_2022,Tolias_JCP_2023}.

In the present work, we shall develop two-dimensional finite-temperature extensions for the prototypical members of the aforementioned two branches of the self-consistent dielectric formalism; the STLS scheme and the HNC scheme. The schemes are both semi-classical in nature with quantum effects entering the density response function exclusively through the non-interacting density response, which implies that the LFC is purely static, i.e., $G_D(\mathbf{q}, i\omega_{\ell}) \rightarrow G_D(\mathbf{q})$. In other words, quantum effects are included at the RPA level.

Regardless of the dimensionality, neither the STLS nor the HNC scheme abide by the compressibility sum rule (CSR)~\cite{quantum_theory}. However, the static LFC of the STLS scheme can be extended to satisfy the CSR~\cite{vs_original}. Despite the large mathematical and numerical complexity caused by this extension, it has been observed that this yields marginal improvements in the structural predictions~\cite{Tolias_PRB_2024}. For the finite-temperature 3D-UEG, this has been attributed to the fact that the enforcement of small wavenumber consistency (satisfaction of the CSR, which is violated by the STLS scheme) has the undesired side effect of large wavenumber inconsistency (violation of the cusp condition, which is satisfied by the STLS scheme)~\cite{Tolias_PRB_2024}.

\subsubsection{STLS and HNC approximations}\label{subsub1}

\noindent The general form of the STLS LFC is~\cite{stls_original}
\begin{equation*}
    G_D^{\text{STLS}}(\mathbf{q}) = -\frac{1}{n} \int \frac{d^Dk}{(2\pi)^D} \frac{U_D(\mathbf{k})}{U_D(\mathbf{q})} \frac{\mathbf{q} \cdot \mathbf{k}}{q^2} [S(\mathbf{q}-\mathbf{k})-1].
\end{equation*}
Regardless of the dielectric scheme, somewhat unexpectedly, obtaining a numerically convenient form for the 2D case is considerably more challenging than in the 3D case. The change of variables $\cos\theta=x$ that simplifies inclination angle integrations in 3D spherical coordinates results in more demanding angular integrations in 2D polar coordinates. Consequently, the derivation of tractable expressions for 2D schemes requires more elaborate strategies. After introducing the normalizations for the wavevector $x=q/q_F$ and the integration variable $y=k/q_F$, we obtain the numerically convenient form
\begin{equation}
\begin{split}
    G_{2D}^{\mathrm{STLS}}(x) 
    &= - \int_{0}^{\infty} dy\, y\, [S(y)-1] \times\\
    &\Bigg[
        \frac{x-y}{\pi x}\,
        \mathcal{K}\!\left(\frac{2\sqrt{xy}}{x+y}\right)
        + 
        \frac{x+y}{\pi x}\,
        \mathcal{E}\!\left(\frac{2\sqrt{xy}}{x+y}\right)
    \Bigg],
\end{split}
\label{eq:STLS_LFC}
\end{equation}
where $\mathcal{K}(m)=\int_0^{\pi/2}(1-m^2\sin^2{\phi})^{-1/2}d\phi$ and $\mathcal{E}(m)=\int_0^{\pi/2}(1-m^2\sin^2{\phi})^{1/2}d\phi$ are the elliptic integrals of the first kind and second kind. The above LFC expression is equivalent to the LFC expression derived by Schweng \& B\"ohm~\cite{Schweng_1994}, but it is numerically evaluated much faster courtesy of the special functions. The essential steps of this derivation are presented in Appendix~\ref{appendix:STLS_HNC}. 

The general form of the HNC LFC is~\cite{tanaka_hnc}
\begin{align*}
    G_D^{\text{HNC}}(\mathbf{q}) =& -\frac{1}{n} \int \frac{d^Dk}{(2\pi)^D}  \frac{U_D(\mathbf{k})}{U_D(\mathbf{q})}\frac{\mathbf{q} \cdot \mathbf{k}}{q^2}\times\\
    &[S(\mathbf{q}-\mathbf{k}) - 1] \{1 - [G(\mathbf{k}) - 1][S(\mathbf{k}) - 1]\}.
\end{align*}
After introducing the normalizations for the wavevector $x=q/q_F$ and the integration variable $y=k/q_F$ we obtain the numerically convenient form 
\begin{equation}
\begin{split}
    &G_{2D}^{\mathrm{HNC}}(x) 
    =\; G_{2D}^{\mathrm{STLS}}(x) 
    + \frac{1}{\pi x} 
    \int_{0}^{\infty} \,du[G(u)-1]\times\\
    &\,\,\,\,[S(u)-1]\int_{|u-x|}^{u+x} w dw
        \frac{\operatorname{sgn}(u^{2}-w^{2}+x^{2})}{\sqrt{\frac{4u^2x^2}{(u^2-w^2+x^2)^2}-1}}
        [S(w)-1]
\end{split}
\label{eq:HNC_LFC}
\end{equation}
where $\mathrm{sgn}(x)=|x|/x$ denotes the sign function. The essential steps of this derivation are again presented in Appendix~\ref{appendix:STLS_HNC}.

\subsubsection{Normalized set of equations\label{subsub2}}

\noindent In view of the arbitrary dimensionality finite temperature dielectric formalism framework outlined in Sec.~\ref{sec:Dielectric}, we now present the computational procedure used to evaluate the properties of the 2DEG at any partially degenerate state point $(r_s,\Theta\neq0)$. (\textbf{a}) The reduced chemical potential is computed analytically in two dimensions, since the normalization condition leads to a one-variable integral that can be analytically evaluated with the emerging expression being solvable with respect to $\bar{\mu}$\,\cite{Schweng_1994}:
\begin{equation}
    \bar{\mu}
    =\ln\!\Big[\exp\!\Big(\frac{1}{\Theta}\Big)-1\Big].
    \label{eq:ChemPot}
\end{equation}
(\textbf{b}) Introduction of the auxiliary function 
$\Phi_{2\mathrm{D}}(\mathbf{q},z)=-(E_{\mathrm{F}}/n)\,\widetilde{\chi}_{2\mathrm{D},0}(\mathbf{q},z)$,
in complex frequency space $z$ which translates to the form
$\widetilde{\chi}_{2\mathrm{D},0}(x,\ell)/(n\beta) = -\Theta\,\Phi_{2\mathrm{D}}(x,\ell)$ in Matsubara space, leads to the computationally convenient representation\,\cite{Schweng_1994}
\begin{equation}
\begin{split}
    \Phi_{2D}&(x,\ell\neq 0)= \int_{0}^{\infty} dy\,\frac{y}{\exp\!\left(\frac{y^{2}}{\Theta}-\bar{\mu}\right)+1}\times\\
&\frac{2|\cos{\phi}|}{\bigg\{
\big[\tfrac{x^{4}}{4}-x^{2}y^{2}-(\pi\ell\Theta)^{2}\big]^{2}
+ x^{4}(\pi\ell\Theta)^{2}
\bigg\}^{1/4}},\label{eq:ideal}\\
\end{split}
\end{equation}
where $\tan(2\phi)= {x^{2}(\pi\ell\Theta)}/{[{x^{4}}/{4}-x^{2}y^{2}-(\pi\ell\Theta)^{2}]}$. When $\tan(2\phi)$ is negative, the substitution $2\phi\rightarrow\pi-2\phi$ must be applied to preserve the correct branch. The zero frequency term $\ell=0$ must be treated separately in order to avoid the removable singularity at $y=x/2$
\begin{equation}
\begin{split}
\Phi_{2D}&(x,0)= 1 - \exp\left(-\frac{1}{\Theta}\right)\\
&- \frac{1}{\Theta x} 
\int_{0}^{x/2} dy\,
\frac{1}{\cosh^{2}\!\left(\frac{y^{2}}{2\Theta}-\frac{\bar{\mu}}{2}\right)}
y\sqrt{\frac{x^2}{4}-y^2}.
\end{split}
\end{equation}
(\textbf{c}) The non-interacting (Hartree-Fock) SSF is computed through
\begin{align}
S&_{\mathrm{HF}}(x)
= \frac{2}{\pi}\int_{0}^{\infty} dy\,
\frac{y}{\exp(\frac{y^{2}}{\Theta}-\bar{\mu})+1}
\int_{0}^{\pi} \Big\{\coth\Big(\frac{x^{2}}{2\Theta}+\frac{xy}{\Theta}\nonumber\\
&\cos{\phi}\Big)-\Big(\frac{x^2}{2\Theta}+\frac{xy}{\Theta}\cos\phi\Big)^{-1}\Big\}d\phi\
+ \Theta\,\Phi_{2D}(x,0).
\end{align}
where we have added and subtracted the static auxiliary density response $-\Theta\Phi(x,0)$ to remove the singularity of the hyperbolic cotangent. (\textbf{d}) The SSF is computed from Eq.(\ref{eq:SSF2}) with the initial condition $G_{\text{2D}}(x)=0$ (RPA)
\begin{equation}
\begin{split}
S(x)&= S_{\mathrm{HF}}(x)
- \sqrt{2}\, r_{s}\,\Theta \frac{1}{x} [1-G_{2D}(x)]\times\\
&\sum_{\ell=-\infty}^{+\infty}
\frac{\Phi^2_{2D}(x,\ell)}{1 + \sqrt{2}\, r_{s}\frac{1}{x}[1-G_{2D}(x)]\Phi_{2D}(x,\ell)} ,
\end{split}
\end{equation}
where we have employed the standard convergence acceleration technique of adding and subtracting $S_{\mathrm{HF}}(x)$\,\cite{stls}. Additional numerical efficiency is obtained by adding and subtracting, within the Matsubara summation, the joint $x\!\to\!\infty$ and $\ell\!\to\!\infty$ asymptotic form of $\Phi_{2\mathrm{D}}(x,\ell)$ which also obeys
\begin{equation*}
    \label{eq:placeholder_label}
    \sum_{\ell=-\infty}^{\infty} \Phi_\infty^2(x,\ell)
    = \frac{1}{2 \Theta^2} \left[ \text{csch}^2\!\left(\frac{x^2}{2\Theta}\right)
    + \frac{2\Theta}{x^2} \coth\!\left(\frac{x^2}{2\Theta}\right) \right].
\end{equation*}
(\textbf{e}) The LFC $G_{2\mathrm{D}}^{\mathrm{STLS}}(x)$ from Eq.~(\ref{eq:STLS_LFC}) or $G_{2\mathrm{D}}^{\mathrm{HNC}}(x)$ from Eq.~(\ref{eq:HNC_LFC}), depending on the chosen closure, are evaluated. (\textbf{f}) The last two steps are iterated until the convergence criterion $\sum_{i}\big| G_{2\mathrm{D}}^{(n)}(x_i)-G_{2\mathrm{D}}^{(n-1)}(x_i) \big| \le 10^{-5}$ is satisfied.

\subsection{Path integral Monte Carlo\label{sec:PIMC}}

\begin{figure}
    \centering
    \includegraphics[width=0.485\textwidth]{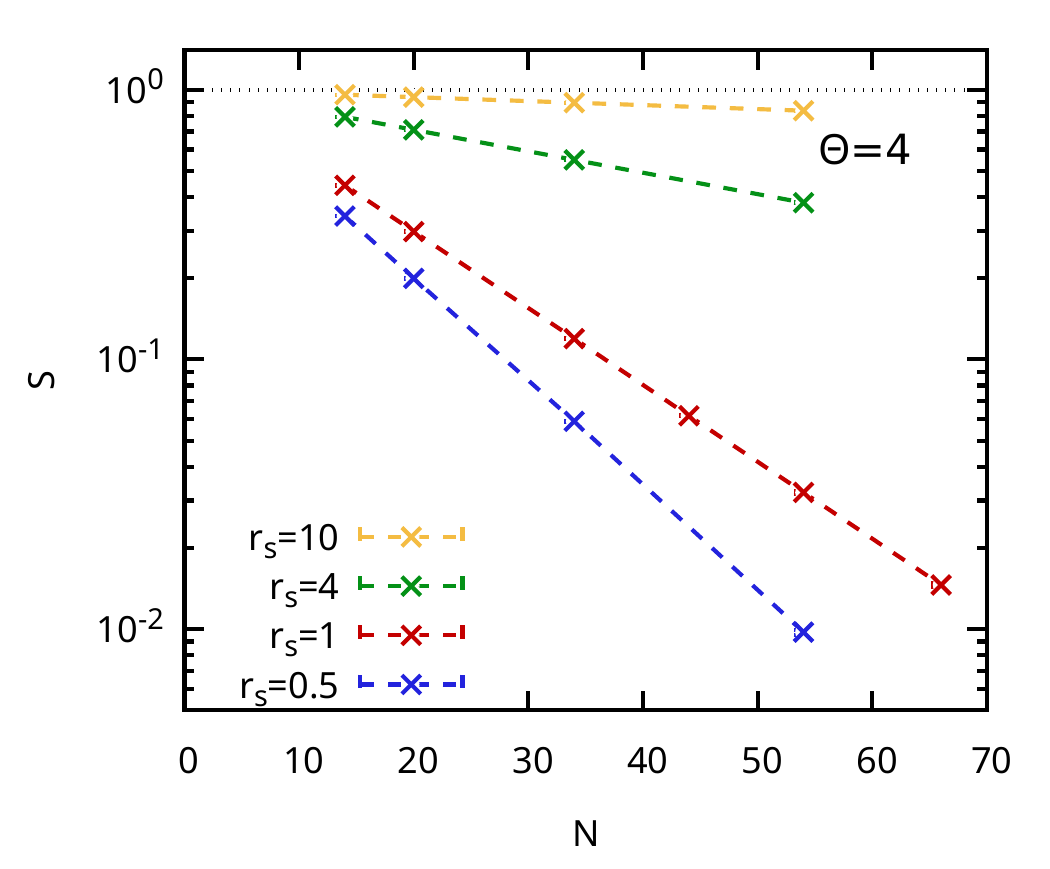}
    \caption{Average sign $S$ of the 2DEG as a function of particle number $N$ at $\Theta=4$ and various values of the coupling parameter $r_s$.
    }
    \label{fig:sign}
\end{figure}

\noindent We use the direct path integral Monte Carlo method to simulate the 2DEG in the canonical ensemble, i.e., the inverse temperature $\beta$, the number density $n=N/L^2$ and the box length $L$ are fixed. Accessible introductions to the PIMC method are available in the literature~\cite{cep,boninsegni1}, thus we limit ourselves to a concise overview of the relevant procedure. We simulate the 2DEG with a finite particle number $N$ and system size, implementing standard periodic boundary conditions. The long-range Coulomb interaction is taken into account using the Ewald sum following the notation by Osychenko \emph{et al.}~\cite{Osychenko20022012}, which we have implemented into the open-source \texttt{ISHTAR} code~\cite{ISHTAR}. We employ the \emph{primitive factorization} $e^{-\epsilon\hat{H}}\approx e^{-\epsilon\hat{K}}e^{-\epsilon\hat{V}}$, where $\epsilon=\beta/P$ with $\hat{K}$, $\hat{V}$ denoting the kinetic and interaction contributions to the full Hamiltonian. The convergence with the number of high-temperature factors $P$ has been carefully checked, see the Appendix~\ref{appendix:PIMC}. We note that higher-order factorizations have been presented in the literature~\cite{Chin_PRE_2015,sakkos_JCP_2009,Dornheim_NJP_2015,Zillich_JCP_2010}, but they are not required here. 

\begin{figure*}
    \centering
    \includegraphics[width=0.325\textwidth]{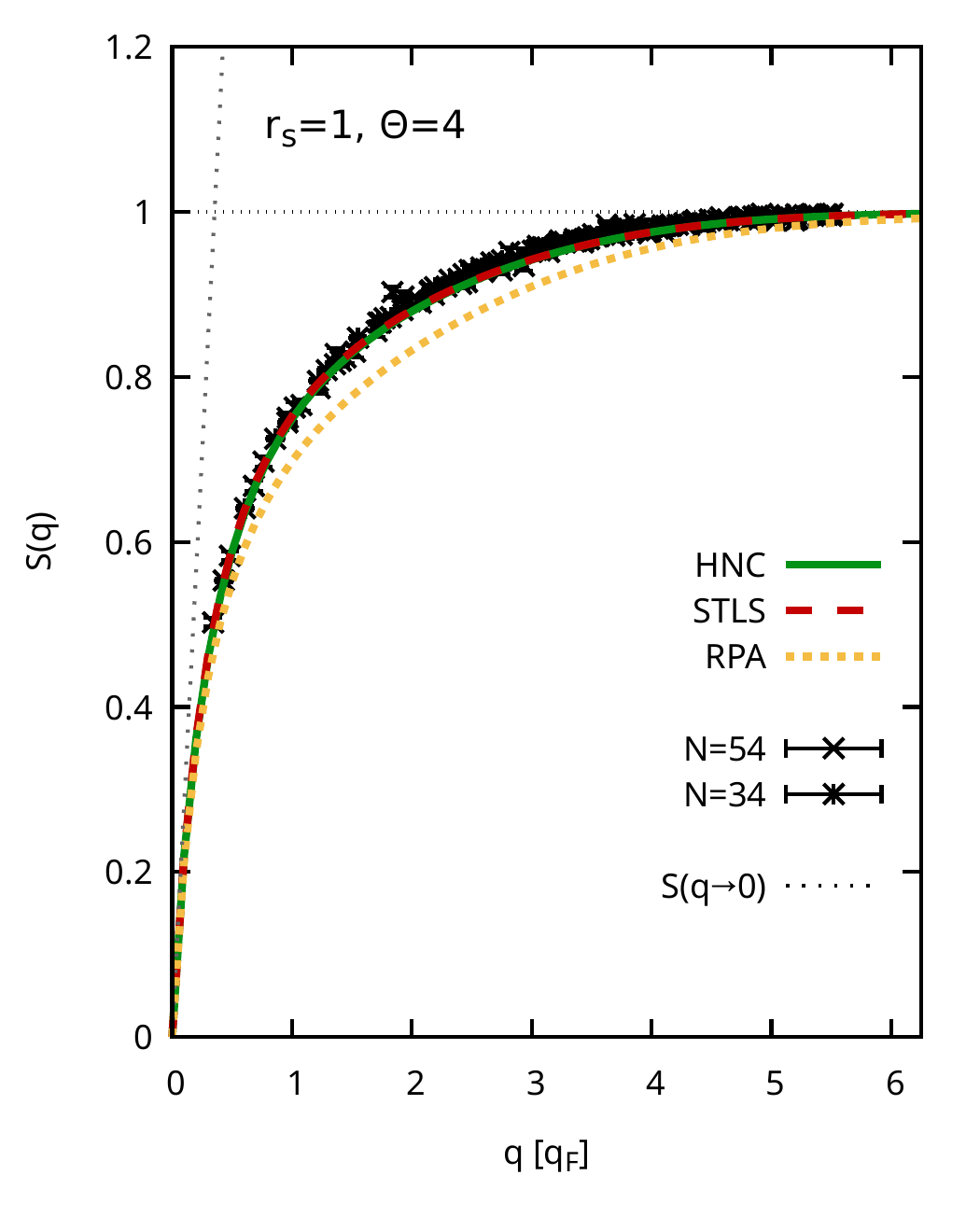}\includegraphics[width=0.325\textwidth]{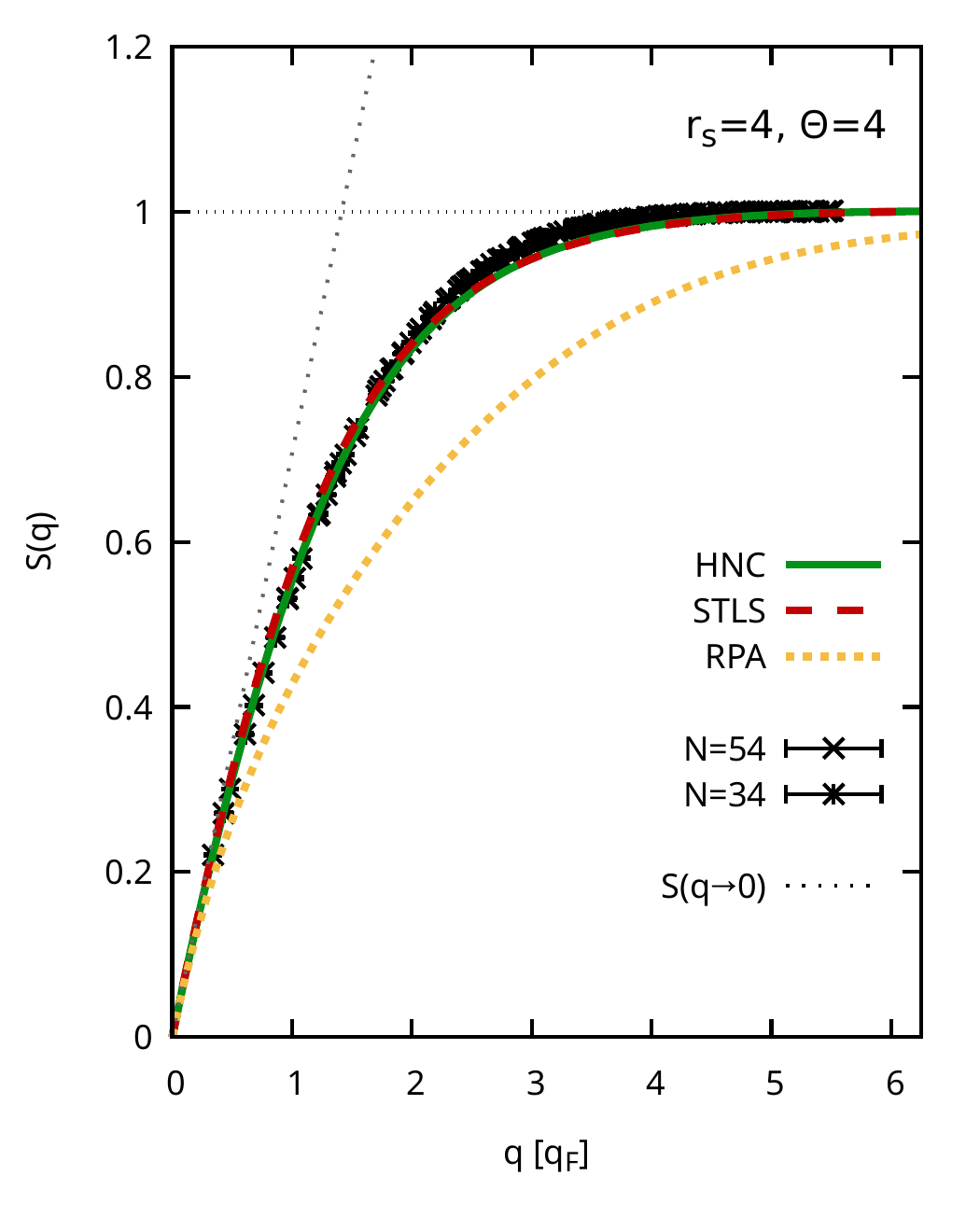}\includegraphics[width=0.325\textwidth]{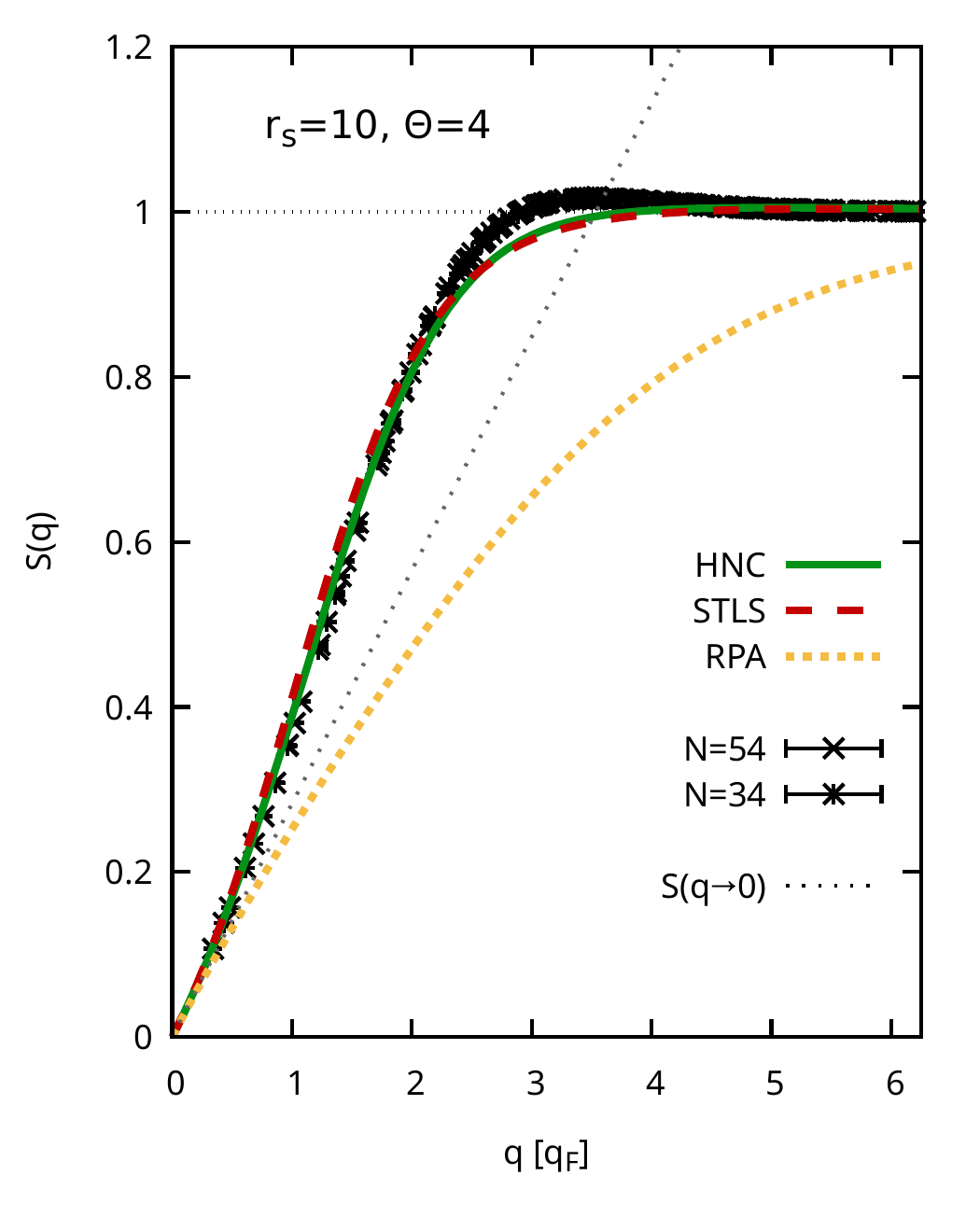}\\\vspace*{-1cm}\includegraphics[width=0.325\textwidth]{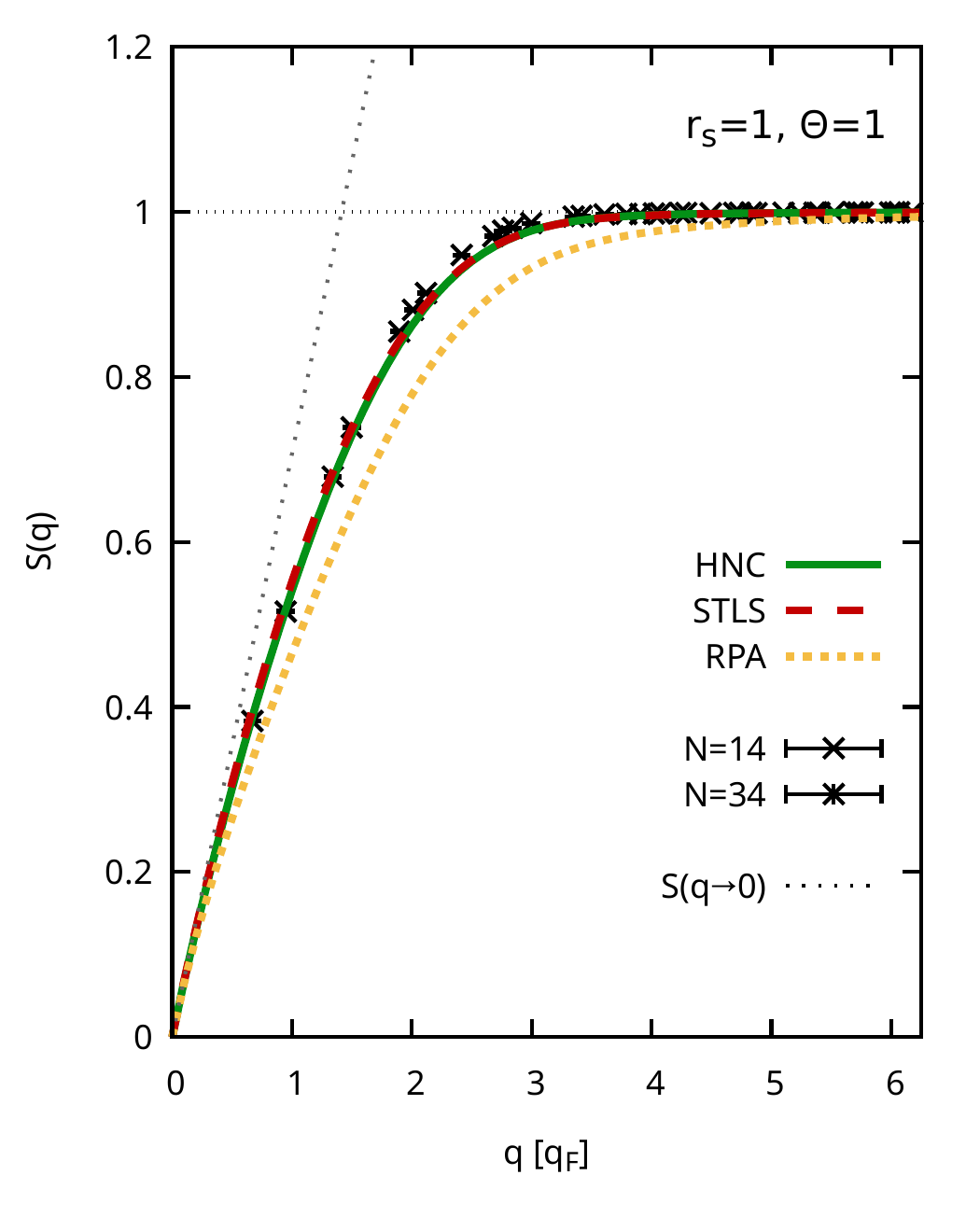}\includegraphics[width=0.325\textwidth]{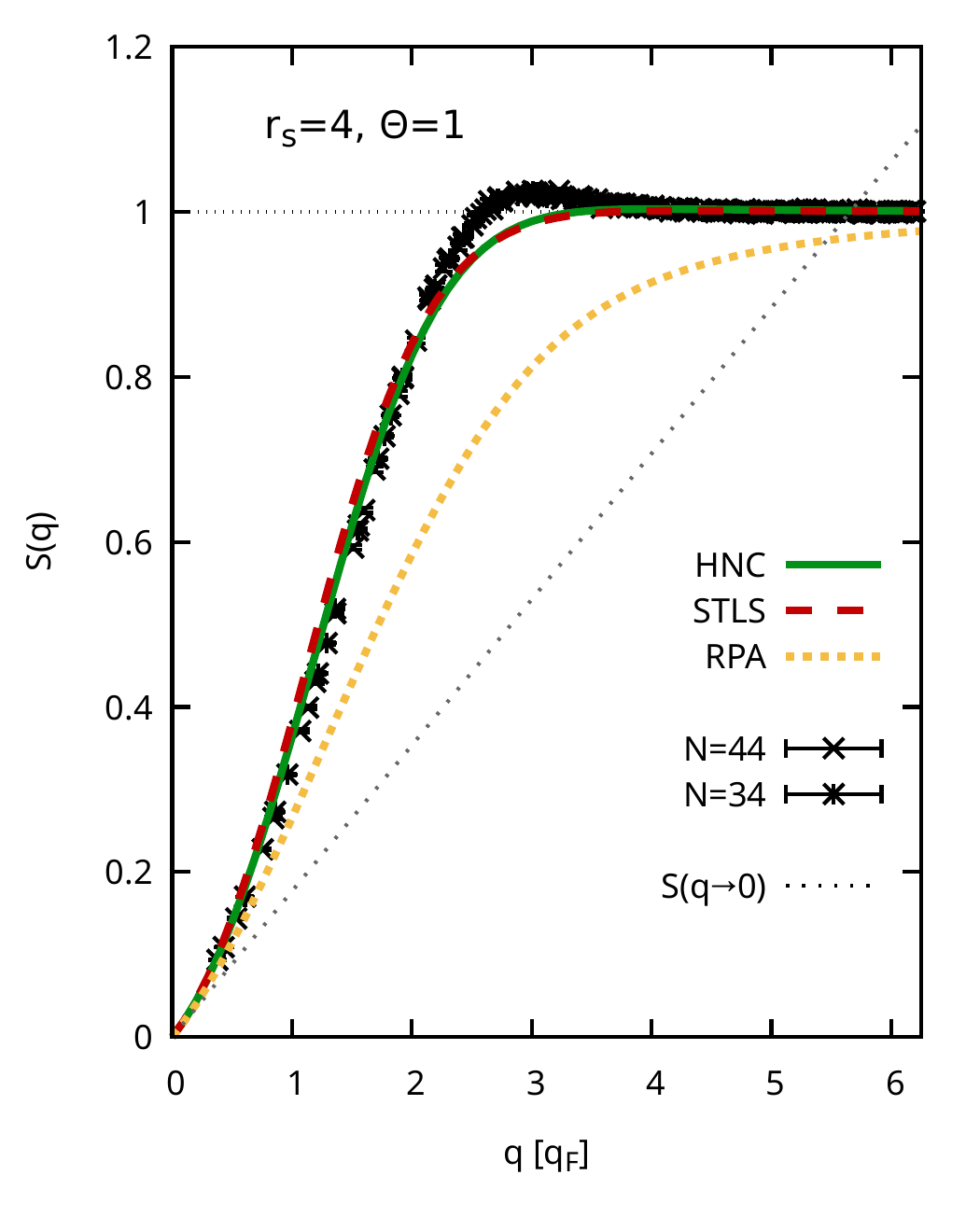}\includegraphics[width=0.325\textwidth]{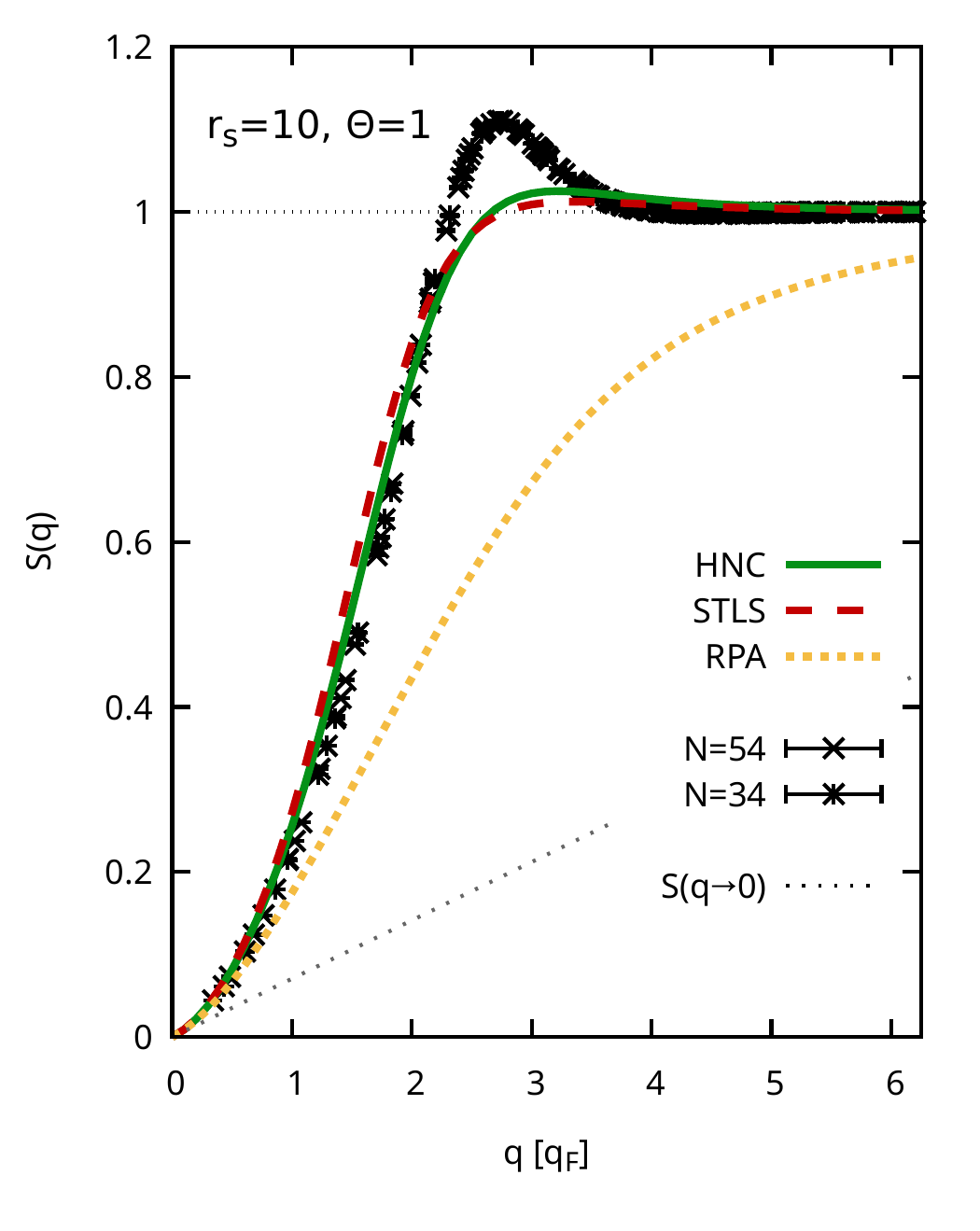}
    \caption{Static structure factor $S(\mathbf{q})$ of the 2DEG at $r_s=1$ (left), $r_s=4$ (center), and $r_s=10$ (right) with the top and bottom rows corresponding to $\Theta=4$ and $\Theta=1$, respectively. Black symbols: quasi-exact PIMC reference data; solid green: HNC scheme; dashed red: STLS scheme; dotted yellow: RPA. The dotted gray lines correspond to the exact $q\to0$ limit, see Eq.~(\ref{eq:SSF0}).
    }
    \label{fig:SSF}
\end{figure*}

\begin{figure*}
    \centering
    \includegraphics[width=0.45\textwidth]{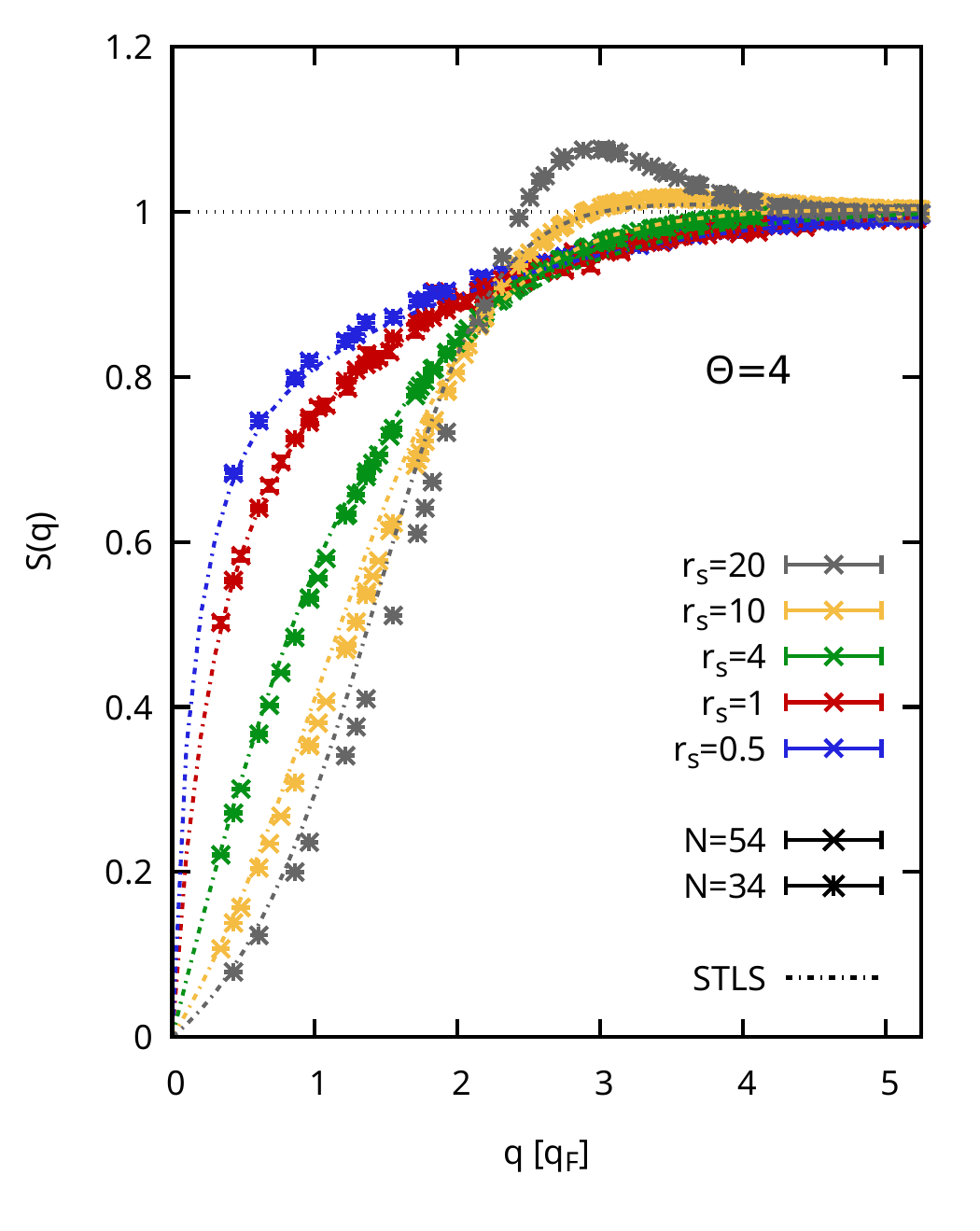}\includegraphics[width=0.45\textwidth]{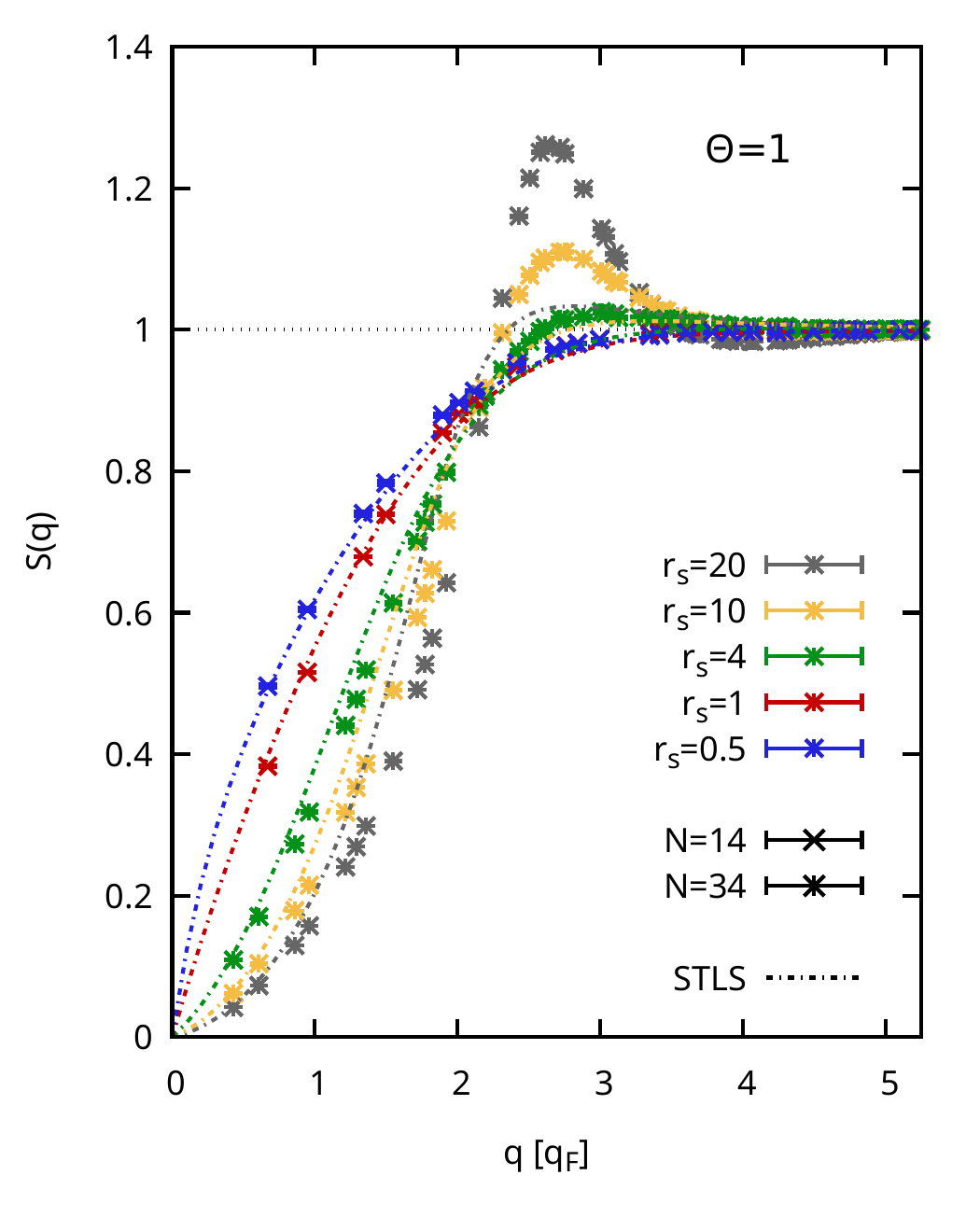}
    \caption{Dependence of the 2DEG static structure factor $S(\mathbf{q})$ on the coupling parameter $r_s$ at $\Theta=4$ (left) and $\Theta=1$ (right). The stars and crosses show quasi-exact PIMC reference results for different particle numbers $N$, while the gray, yellow, green, red, and blue symbols correspond to $r_s=20, 10, 5, 1, 0.5$, respectively. The STLS results are included as dash-dotted lines with the same color code. 
    }
    \label{fig:SSF_rs}
\end{figure*}

In order to consider the required anti-symmetry of the thermal density matrix under the exchange of the coordinates of electrons of the same spin-orientation, we have to sample all possible permutation cycles in PIMC~\cite{Dornheim_permutation_cycles}. This is achieved efficiently by using a canonical adaptation of the worm algorithm by Boninsegni \emph{et al.}~\cite{boninsegni1,boninsegni2}, see Ref.~\cite{Dornheim_PRB_nk_2021} for details. The anti-symmetry further constitutes the root cause of the notorious fermion sign problem~\cite{troyer,dornheim_sign_problem,Loh_PRB_1990}, which leads to an exponential increase in the required compute time with respect to key system parameters such as $N$ and $\beta$. This is conveniently quantified by the average sign $S=Z_\textnormal{Fermi}(\beta,n,L)/Z_\textnormal{Bose}(\beta,n,L)$, which constitutes a direct measure for the degree of cancellation of positive and negative terms in the fermionic partition function $Z_\textnormal{Fermi}(\beta,n,L)$. In Fig.~\ref{fig:sign}, we show the average sign from our PIMC simulations as a function of the system size at $\Theta=4$ and various values of the density parameter $r_s$.
First, we find a pronounced dependence of our results on $r_s$, which would be absent for non-interacting Fermi systems at the same conditions. In fact, the UEG becomes more strongly coupled for lower density~\cite{Ott2018,quantum_theory}, and the corresponding Coulomb repulsion suppresses the formation of permutation cycles~\cite{Dornheim_permutation_cycles}, which are the origin of the sign problem. Second, we observe the expected exponential decrease of $S$ with $N$. In practice, PIMC simulations remain feasible for $S\gtrsim0.01$. We stress that we do not employ any nodal restrictions here~\cite{Ceperley1991}. Hence, our simulations are computationally involved, but exact within the given Monte Carlo error bars, i.e., "quasi-exact". For completeness, we note that a variety of alternative strategies to deal with the sign problem have been presented in the literature~\cite{Hirshberg_JCP_2020,Dornheim_JCP_2020,Dornheim_NJP_2015,Dornheim_CPP_2019,Chin_PRE_2015,Malone_JCP_2015,Joonho_JCP_2021,Xiong_JCP_2022,Xiong_JCP_2025,Dornheim_JCP_xi_2023,Dornheim_NatComm_2025,dornheim2025taylorseriesperspectiveab,Filinov_PRE_2020}, but they will not be considered here.

\section{Results}

\noindent An online repository with all presented PIMC results is freely available online~\cite{repo}.

\subsection{Static structure factor\label{sec:SSF}}

\noindent Let us begin our investigation with the static structure factor $S(\mathbf{q})$.
In Fig.~\ref{fig:SSF}, we compare our STLS (dashed red), HNC (solid green) as well as RPA (dotted yellow) results to quasi-exact PIMC simulations (black symbols) for $\Theta=4$ (top row) and $\Theta=1$ (bottom row) over a broad range of densities. \emph{First}, we find no differences between PIMC results for different system size (crosses and stars) for all considered parameters except for the different $\mathbf{q}$-grid discretization. This is consistent with previous PIMC based investigations of the 3D UEG at comparable densities and temperatures~\cite{review,Dornheim_JCP_2021}. \emph{Second}, we find that the RPA, which describes the dynamic density response on the mean-field level, exhibits considerable inaccuracies for intermediate wavenumbers $q$ even at $r_s=1$ and, thus, should not be utilized for practical applications except possibly in the regime of extremely high densities. It only becomes accurate for $q\gg q_\textnormal{F}$, i.e., in the single-particle regime where correlation effects vanish by definition, and for $q\to0$, i.e., in the collective long-wavelength regime where the density correlations are determined by the undamped plasmon excitations; which manifest themselves in the exact asymptotic
\begin{eqnarray}\label{eq:SSF0}
    S(q\to0) = \frac{q}{2\pi n \beta}\ ,
\end{eqnarray}
see the light dashed gray lines in Fig.~\ref{fig:SSF}. We highlight the linear screening in the 2D case, see Eq.~(\ref{eq:SSF0}), in contrast to the well known parabolic screening in the 3D case~\cite{kugler_bounds}. 
\begin{equation*}
    S(q\to0) = \frac{\hbar{q}^2}{2m\omega_{\mathrm{pl}}}\coth{\left(\frac{1}{2}\beta\hbar\omega_{\mathrm{pl}}\right)}\,,
\end{equation*}
where $\omega_{\mathrm{pl}}$ is the 3D electron plasma frequency. This difference is a consequence of the dependence of the functional form of the plasmon dispersion relation on the system dimensionality~\cite{quantum_theory}. \emph{Third}, we observe a pronounced systematic improvement of the STLS, HNC schemes over the RPA with respect to the PIMC data. At $\Theta=4$, STLS and HNC give very similar results for all three considered values of $r_s$, with HNC only showing a clear improvement over the former for $r_s=10$, $q\lesssim2q_\textnormal{F}$. At $\Theta=1$, the differences are somewhat more pronounced, with HNC being the more accurate choice for the description of $S(\mathbf{q})$.

\begin{figure*}
    \centering
    \includegraphics[width=0.45\textwidth]{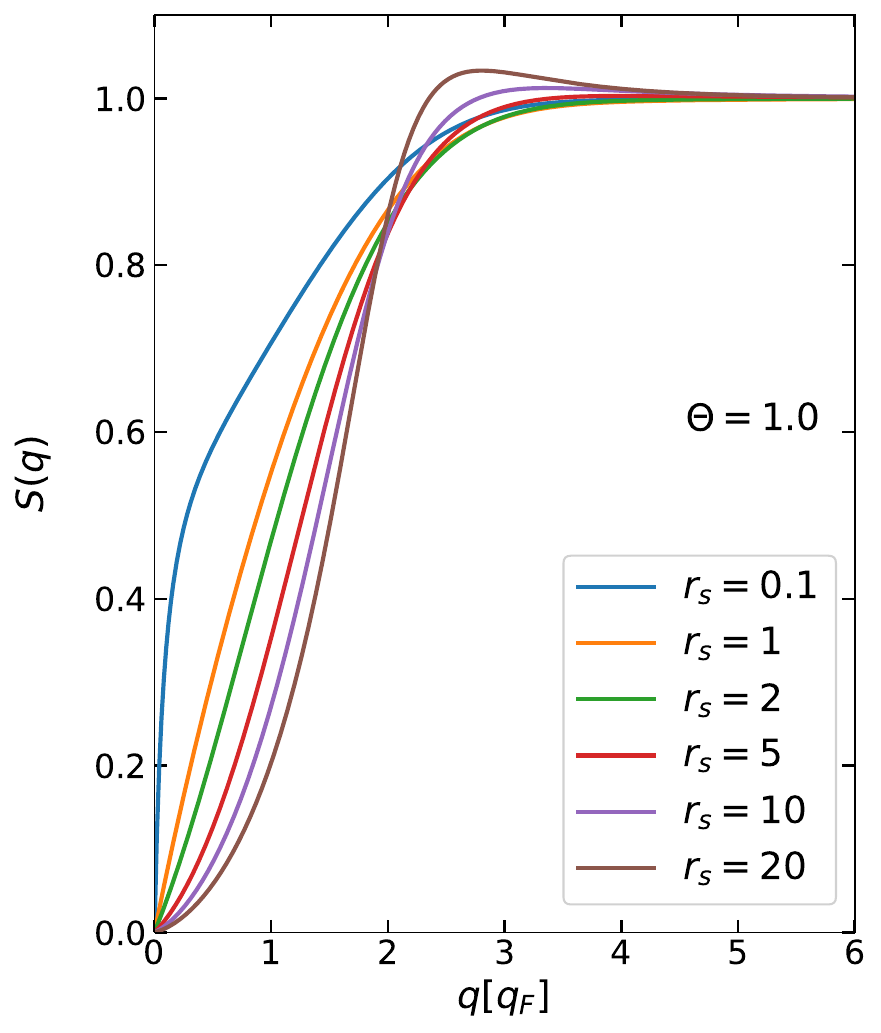}\includegraphics[width=0.45\textwidth]{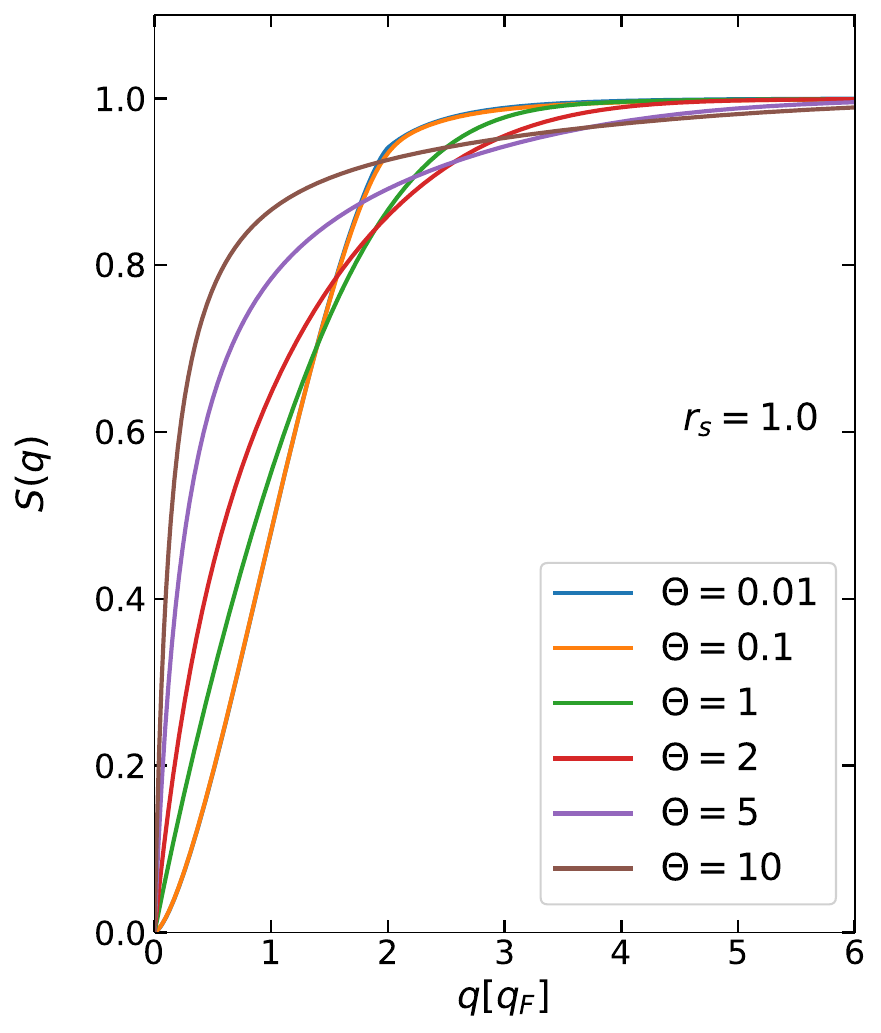}
    \caption{Parametric sweep for the STLS generated 2DEG static structure factor $S(\mathbf{q})$. Left: dependence of $S(\mathbf{q})$ on the coupling parameter $r_s=0.1,1,2,5,10,20$ at $\Theta=1.0$. Right: dependence of $S(\mathbf{q})$ on the degeneracy parameter
    $\Theta=0.01,0.1,1,2,5,10$ at $r_s=1.0$. 
    }
    \label{fig:SSF_stls_rs_theta_sweep}
\end{figure*}

However, the interaction energy is obtained by an integration over $S(\mathbf{q})-1$, see Eq.~(\ref{eq:uint}) that follows, whereas the exchange--correlation (XC) free energy requires an additional coupling--constant integration over the interaction energy, see Eq.~(\ref{eq:fxc}) that follows. Consequently, the $S(\mathbf{q})$ deviations in different $q$ regions can partially compensate each other. In particular, the STLS scheme benefits from a particularly favorable cancellation of errors in the integrated SSF, yielding highly accurate thermodynamic quantities despite visible deviations in $S(\mathbf{q})$ at specific wave numbers. In contrast, the improvements of the HNC scheme at small $q$ are only partially offset by the deviations at intermediate and large wavenumbers, resulting in a weaker overall error cancellation. This 2DEG behavior is consistent with earlier findings for the 3D UEG, where STLS was shown to provide remarkably accurate interaction energies and XC free energies in spite of its approximate nature~\cite{review}. Motivated by this observation, we shall focus on the STLS scheme for the computation and parametrization of interaction energies and exchange--correlation free energies in Sec.~\ref{sec:uintfxc}.

In Fig.~\ref{fig:SSF_rs}, we focus more explicitly on the dependence of $S(\mathbf{q})$
on the coupling strength, with the left and right panels corresponding to $\Theta=4$ and $\Theta=1$.
For both temperatures, we observe an SSF decrease with $r_s$ for large wavelengths due to the weaker screening at high density, see also Eq.~(\ref{eq:SSF0}). This trend is reproduced equally well by PIMC and STLS for all cases, as expected. In addition, we find the emergence of a pronounced correlation maximum in $S(\mathbf{q})$ around $q\sim2.5q_\textnormal{F}$ with decreasing density, i.e., increasing coupling strength, in our PIMC results. Specifically, we find a maximum exceeding $1.2$ for $\Theta=1$ and $r_s=20$, indicating the onset of liquid-like behavior~\cite{Kundu_POP_2014}. Evidently, STLS (and also HNC, see Fig.~\ref{fig:SSF} above) fails to capture these coupling effects accurately for $r_s\gtrsim10$. This breakdown occurs for smaller $r_s$ than in the 3D case~\cite{dornheim_electron_liquid,Tolias_JCP_2021,Tolias_JCP_2023,tanaka_hnc}, highlighting the stronger coupling in the 2DEG at equal Wigner-Seitz radius.

In Fig.~\ref{fig:SSF_stls_rs_theta_sweep}, we present the SSF $S(\mathbf{q})$ obtained from the STLS scheme and focus more explicitly on its systematic dependence on the coupling parameter and the degeneracy parameter. The left panel shows $S(\mathbf{q})$ at the fixed reduced temperature $\Theta=1$ and for densities ranging from $r_s=0.1$ to $r_s=20$. As $r_s$ increases, we observe a progressive suppression of $S(\mathbf{q})$ at small wave numbers due to enhanced screening, accompanied by the emergence and growth of a correlation-induced maximum at intermediate $q$, signaling the increasing importance of Coulomb correlations. The right panel displays $S(\mathbf{q})$ at the fixed coupling strength $r_s=1$ and for reduced temperatures ranging from $\Theta=0.01$ to $\Theta=10$. At this moderate coupling strength, the system remains in a weakly correlated regime, and no pronounced correlation peak emerges even at the lowest reduced temperatures. In particular, the nearly identical curves for $\Theta=0.01$ and $\Theta=0.1$ indicate that exchange and correlation effects have essentially saturated and are no longer sensitive to further reductions in temperature. Overall, decreasing $\Theta$ leads to a stronger suppression of $S(\mathbf{q})$ at small $q$, while at intermediate wave numbers the low-temperature curves more rapidly approach the asymptotic limit $S(\mathbf{q})\to1$. Conversely, increasing $\Theta$ weakens correlation effects, resulting in a faster initial rise of $S(\mathbf{q})$ with $q$ but a slower convergence towards unity at large wave numbers.

\begin{figure}
    \centering
    \includegraphics[width=0.44\textwidth]{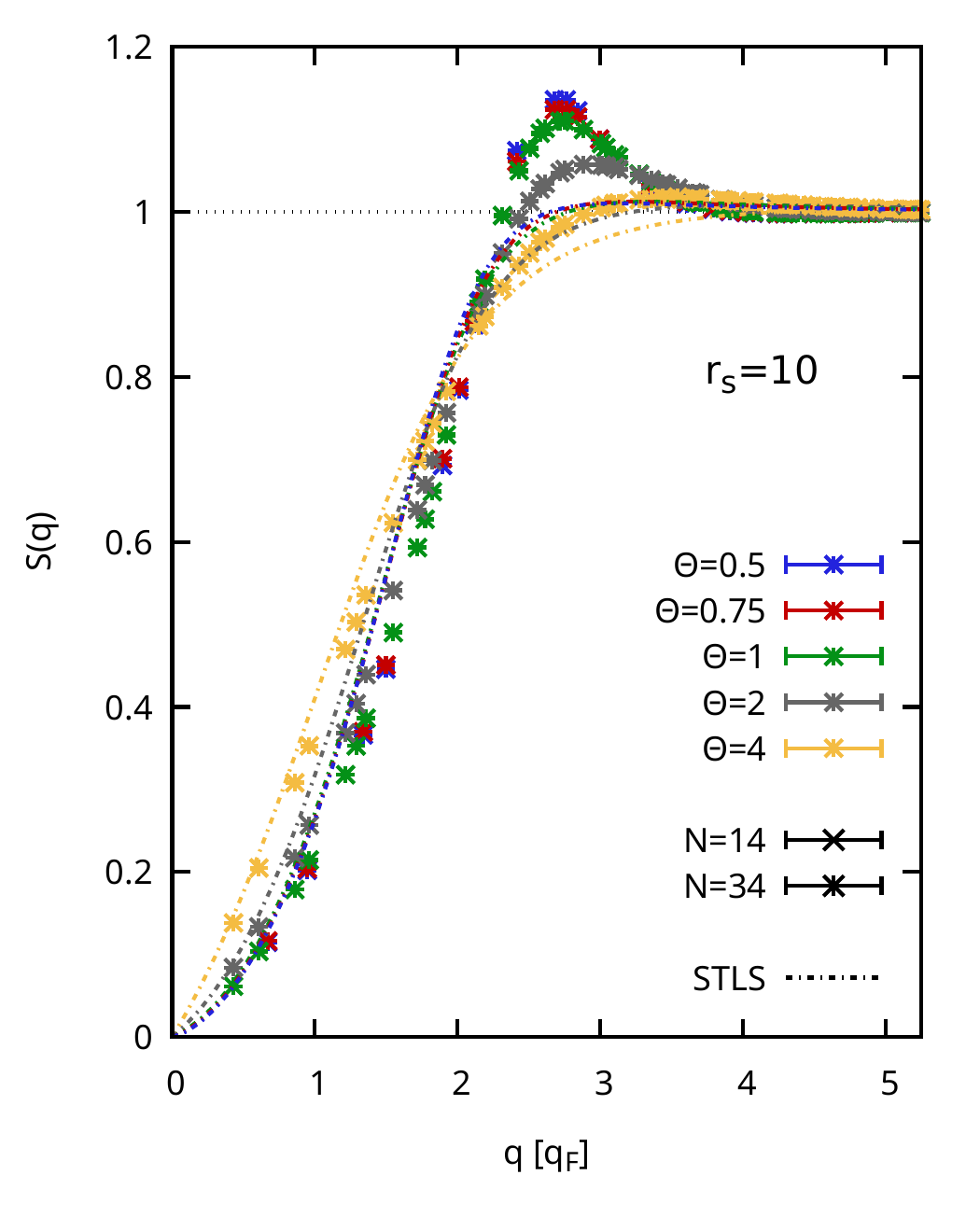}
    \caption{Dependence of the 2DEG static structure factor $S(\mathbf{q})$ on the degeneracy temperature $\Theta=0.5,\,0.75,\,1.0,\,2.0,\,4.0$ at $r_s=10$. The colored symbols and lines correspond to the PIMC data (for $N=14,\,34$) and STLS results, respectively.}
    \label{fig:SSF_temperature}
\end{figure}

In Fig.~\ref{fig:SSF_temperature}, we present the SSF $S(\mathbf{q})$ at a fixed coupling strength $r_s=10$ for various reduced temperatures $\Theta=0.5$, $0.75$, $1$, $2$, $4$ and we compare the STLS scheme results to the PIMC reference data. Overall, we observe a weak dependence of the SSF on temperature in the low- to intermediate-temperature regime. In particular, for $\Theta=0.5$, $0.75$, and $1$, both the magnitude and shape of $S(\mathbf{q})$ are nearly indistinguishable over the entire range of wave numbers. This behavior reflects the dominance of Coulomb-induced spatial correlations at strong coupling, which largely determine the structure of the system and render it relatively insensitive to moderate changes in the temperature. Only at higher temperatures, $\Theta\gtrsim2$, do thermal effects become sufficiently strong leading to a gradual smoothing of $S(\mathbf{q})$ and a reduction of correlation-induced features. The STLS reproduces this weak temperature dependence qualitatively; however, consistent with our earlier observations, it tends to overestimate $S(\mathbf{q})$ at small wave numbers and underestimate it at intermediate $q$, resulting in a very shallow correlation peak.

\begin{figure}
    \centering
    \includegraphics[width=0.495\textwidth]{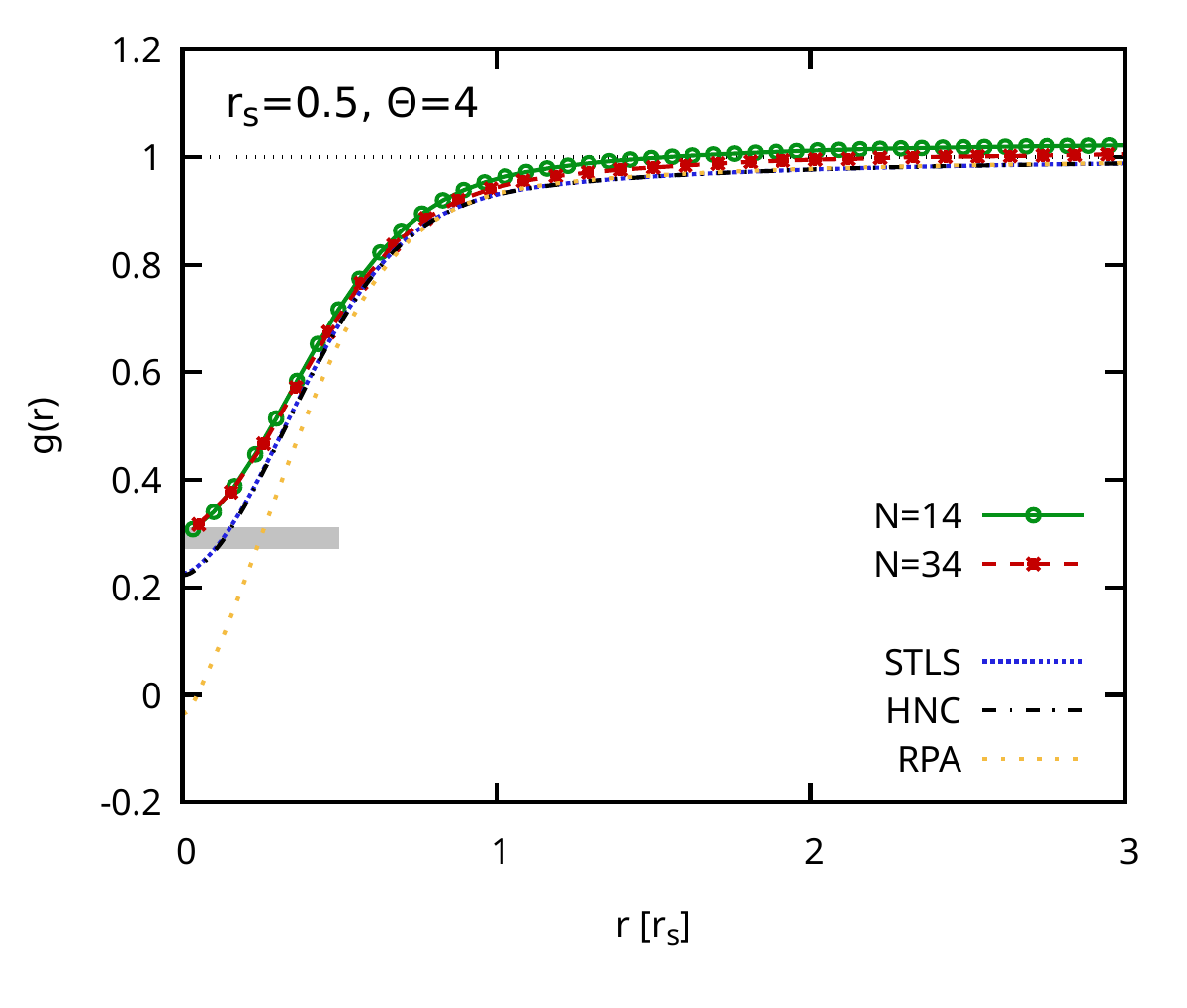}\vspace{-5.1mm}
    \includegraphics[width=0.495\textwidth]{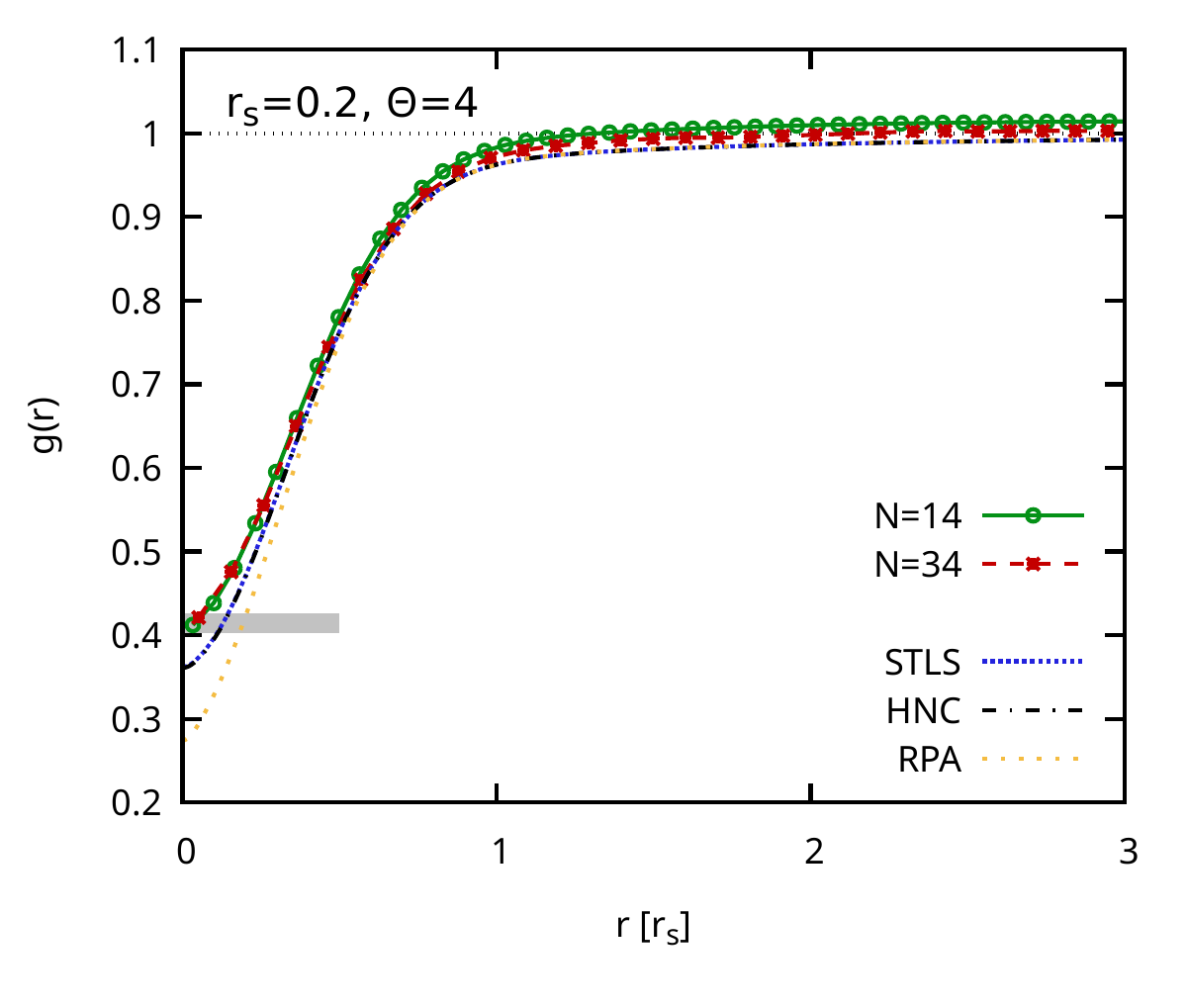}\vspace{-5.1mm}
    \includegraphics[width=0.495\textwidth]{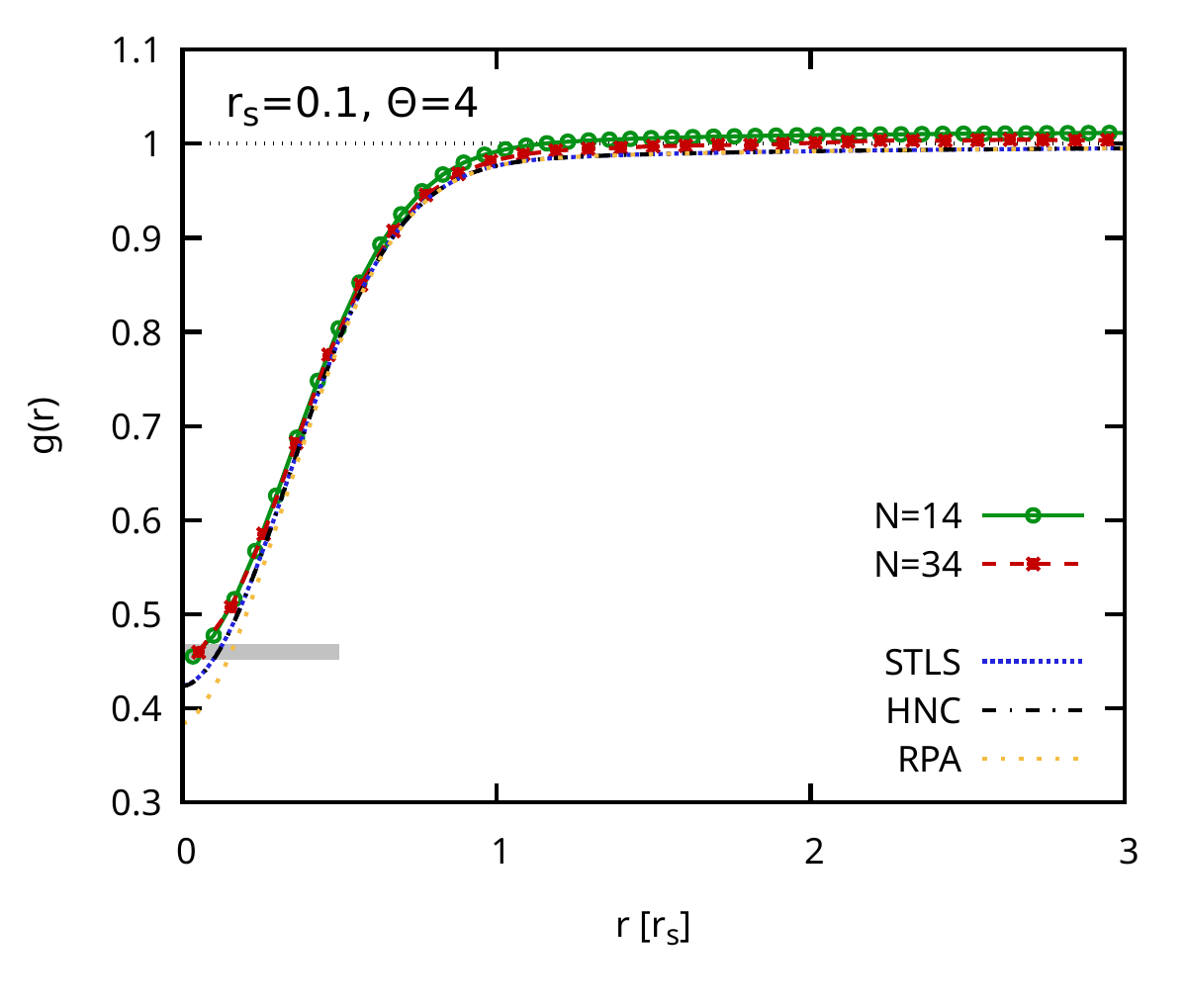}\vspace{-7.5mm}
    \caption{Pair correlation function $g(r)$ of the 2DEG at $\Theta=4$ and $r_s=0.5$ (top), $r_s=0.2$ (mid), $r_s=0.1$ (bottom). Dashed blue: STLS; dash-dotted black: HNC; dotted yellow: RPA; red crosses (green circles): PIMC data for $N=34$ ($N=14$). The shaded gray areas show independent configuration PIMC results for $g(0)$ at $N=14$ and their $2\sigma$-uncertainty interval.}
    \label{fig:Gr_theta4}
\end{figure}

\begin{figure}
    \centering
    \includegraphics[width=0.495\textwidth]{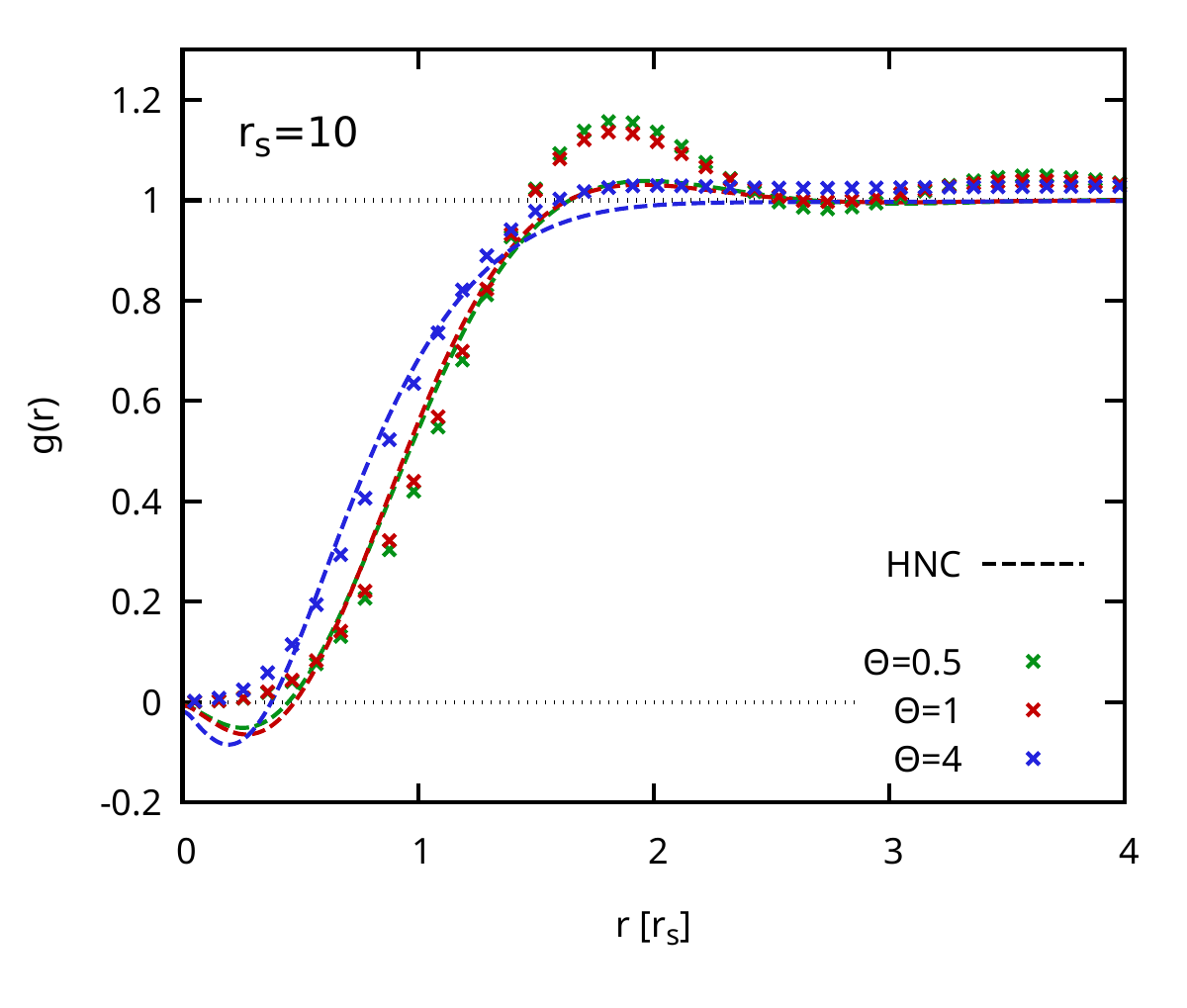}
    \caption{Pair correlation function $g(r)$ of the 2DEG at $r_s=10$ for $\Theta=0.5$ (green), $\Theta=1$ (red), and $\Theta=4$ (blue). Crosses and lines show PIMC reference results for $N=34$ electrons and HNC, respectively.}
    \label{fig:Gr_rs10}
\end{figure}

\begin{figure*}
    \centering
    \includegraphics[width=0.45\textwidth]{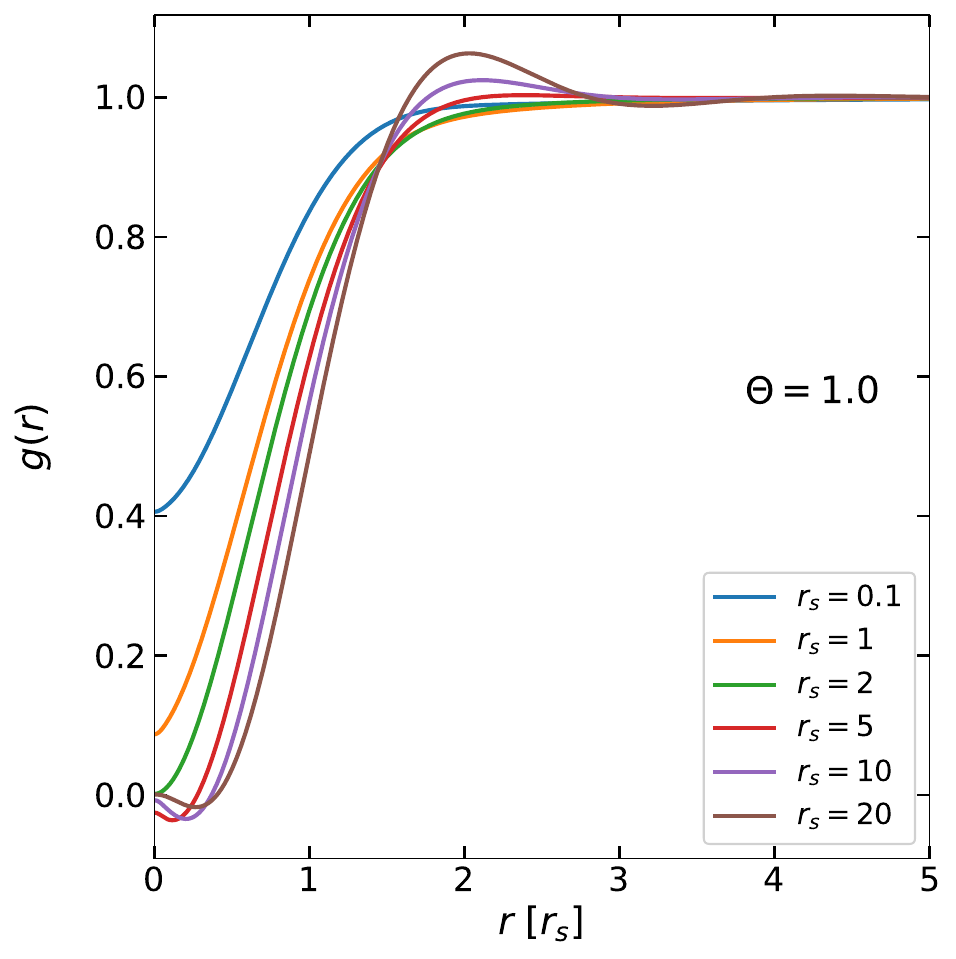}\includegraphics[width=0.45\textwidth]{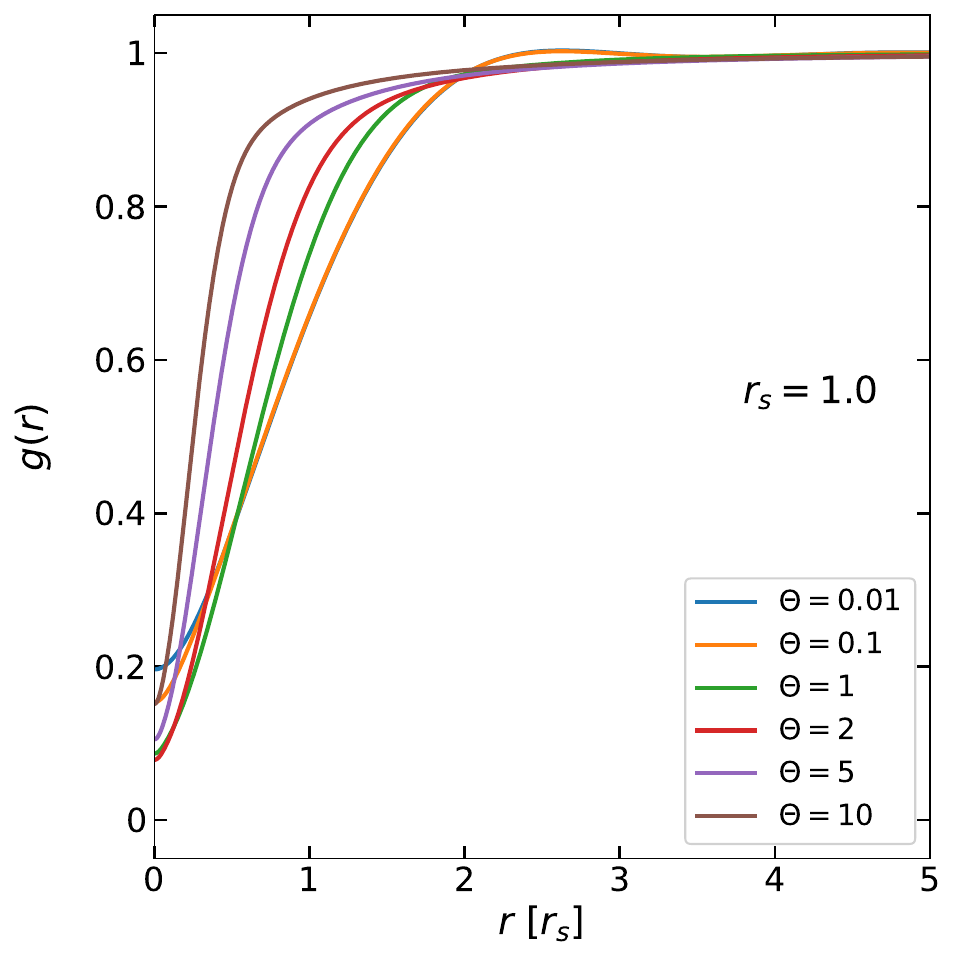}
    \caption{Parametric sweep for the STLS generated 2DEG pair correlation function $g(r)$. Left: dependence of $g(r)$ on the coupling parameter $r_s=0.1,1,2,5,10,20$ at $\Theta=1.0$. Right: dependence of $g(r)$ on the degeneracy parameter
    $\Theta=0.01,0.1,1,2,5,10$ at $r_s=1.0$. 
    }
    \label{fig:pcf_stls_rs_theta_sweep}
\end{figure*}

\subsection{Pair correlation function\label{sec:pcf}}

\noindent We continue our investigation with the pair correlation function $g(r)$, which is the real-space equivalent of the static structure factor. In Fig.~\ref{fig:Gr_theta4}, we use our PIMC reference data to assess the quality of the various considered dielectric schemes at $\Theta=4$ in the limit of relatively high densities. First, in order to cross check our direct PIMC implementation, we carried out independent configuration PIMC (CPIMC) simulations using the openly available \texttt{cpimc.jl} code~\cite{cpimc.jl} that determined the on-top value $g(r=0)$ for $N=14$. Together with the associated uncertainties, the CPIMC on-top values were included as the shaded gray intervals~\cite{cpimc_note}. We find excellent agreement between direct PIMC and CPIMC for all three depicted cases, as expected. Second, to quantify finite size effects, the green circles and red crosses were obtained for $N=14$ and $N=34$, respectively. We find no significant N-dependency for small separations $r\to0$, whereas the large $r$ limits somewhat deviate from each other. Third, some comments are due on the coupling parameter dependence. The top panel of Fig.~\ref{fig:Gr_theta4} corresponds to $r_s=0.5$, which is considered as relatively weak coupling in the context of warm dense matter research~\cite{vorberger2025roadmapwarmdensematter}, but as medium coupling in the context of Green's function theory~\cite{review,Schoof_PRL_2015}. Consequently, the RPA (dotted yellow) is rather inaccurate and even gets slightly negative in the limit of $r\to0$. In contrast, the STLS and HNC results, which cannot be distinguished from each other with the naked eye, exhibit a much better agreement with PIMC, although systematic inaccuracies remain for very small separations. For $r_s=0.2$ (mid panel), coupling effects are significantly less important, and all three schemes perform better than for $r_s=0.5$, even though the RPA remains by far the least accurate. Interestingly, even at $r_s=0.1$ (bottom panel), often considered as weakly coupled in the 3D-UEG literature~\cite{review}, the STLS and HNC results still show systematic errors of the order of $\sim10\%$ at contact.

In Fig.~\ref{fig:Gr_rs10}, we use our PIMC reference data to assess the pair correlation function $g(r)$ performance of the HNC scheme at moderate coupling. Particularly, we plot $g(r)$ results for $r_s=10$ and different temperatures: $\Theta=0.5$ (green), $\Theta=1$ (red), $\Theta=4$ (blue), with the crosses and dashed lines showing the PIMC and HNC results, respectively. First, we find very minor differences between the two lowest temperatures in both HNC and PIMC, consistent with the previously shown results for the static structure factor $S(\mathbf{q})$. Second, our PIMC results exhibit a substantially depleted exchange--correlation hole around $r=0$, which was absent in the high-density regime, see Fig.~\ref{fig:Gr_theta4}. The HNC results substantially overestimate this hard-core region where $g(r)\simeq0$ yielding an unphysical negative pair correlation function that even features a minimum at small but finite $r$. Third, the PIMC results exhibit a coordination shell structure typical of liquids, where a pronounced first peak is followed by a shallow minimum that is followed by an even smaller yet distinguishable second peak. The HNC results have a first peak at the correct position, but with a substantially underestimated height, which is followed by a very shallow minimum that can hardly be resolved with the naked eye.

Finally, in Fig.~\ref{fig:pcf_stls_rs_theta_sweep}, we present the $g(r)$ obtained from the STLS scheme and focus more explicitly on its systematic dependence on the coupling parameter and the degeneracy parameter. The left panel shows $g(r)$ at the fixed
reduced temperature $\Theta= 1$ and for densities ranging from $r_s = 0.1$ to $r_s = 20$. It is evident that the STLS generated $g(r)$ cannot form a exchange-correlation hole and exhibits unphysical behavior. In particular, starting from the non-interacting limit and as the coupling parameter increases: (i) the $g(r)$ near contact is monotonic and positive, (ii) the $g(r)$ near contact remains monotonic but acquires a negative on-top value, (iii) the $g(r)$ near contact becomes non-monotonic with a negative minimum and a negative on-top value, (iv) the $g(r)$ near contact becomes non-monotonic with a negative minimum and a positive on-top value. It is emphasized that the negative $g(r)$ values near contact at moderate-to-strong coupling constitute a well-known pathological feature not only of the STLS scheme but also of all known dielectric schemes (semi-classical or fully quantum)~\cite{Tolias_JCP_2021,Tolias_JCP_2023,Tolias_PRB_2024,Tolias_2025}. Mathematically, the $g(r)$ positivity cannot be enforced in the dielectric formalism, which is based on a self-consistent loop that only guarantees a positive definite static structure factor. On the other hand, the classical-quantum mapping approach, being integral equation theory based, guarantees a pair correlation function positivity~\cite{perrot,Perrot_2DEG_2001}. The right panel shows $g(r)$ at the fixed coupling $r_s=1$ and at temperatures ranging from $\Theta= 0.01$ to $\Theta=10$. It is worth pointing out the complex dependence of the on-top value on $\Theta$, that is monotonically decreasing from $\Theta=0.01$ to $\Theta=1$, nearly constant from $\Theta=1$ to $\Theta=2$ and monotonically increasing from $\Theta=2$ to $\Theta=10$, which is a clear manifestation of the competition between quantum degeneracy and Coulomb correlations. We also observe a weak dependence of $g(r)$ on the reduced temperature at the strongly degenerate regime. This is expected given the weak dependence of $S(q)$ on the temperature at the strongly degenerate regime and the Fourier transform connection between the static structure factor and the pair correlation function. In particular, for $\Theta=0.01$ and $\Theta=0.1$, the $g(r)$ curves are nearly indistinguishable from each other everywhere except from the contact vicinity. 

\begin{figure*}
    \centering
    \includegraphics[width=0.315\textwidth]{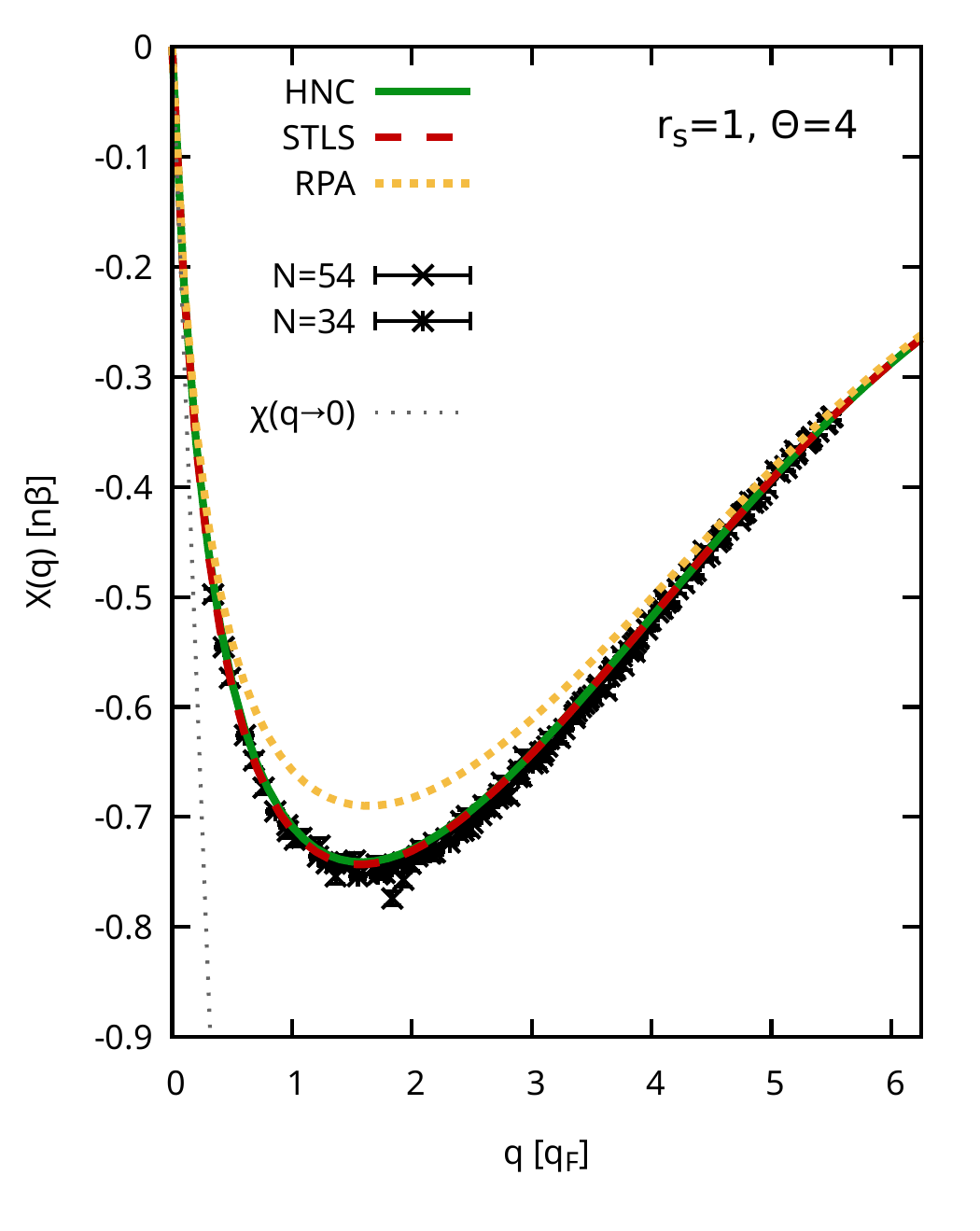}\includegraphics[width=0.315\textwidth]{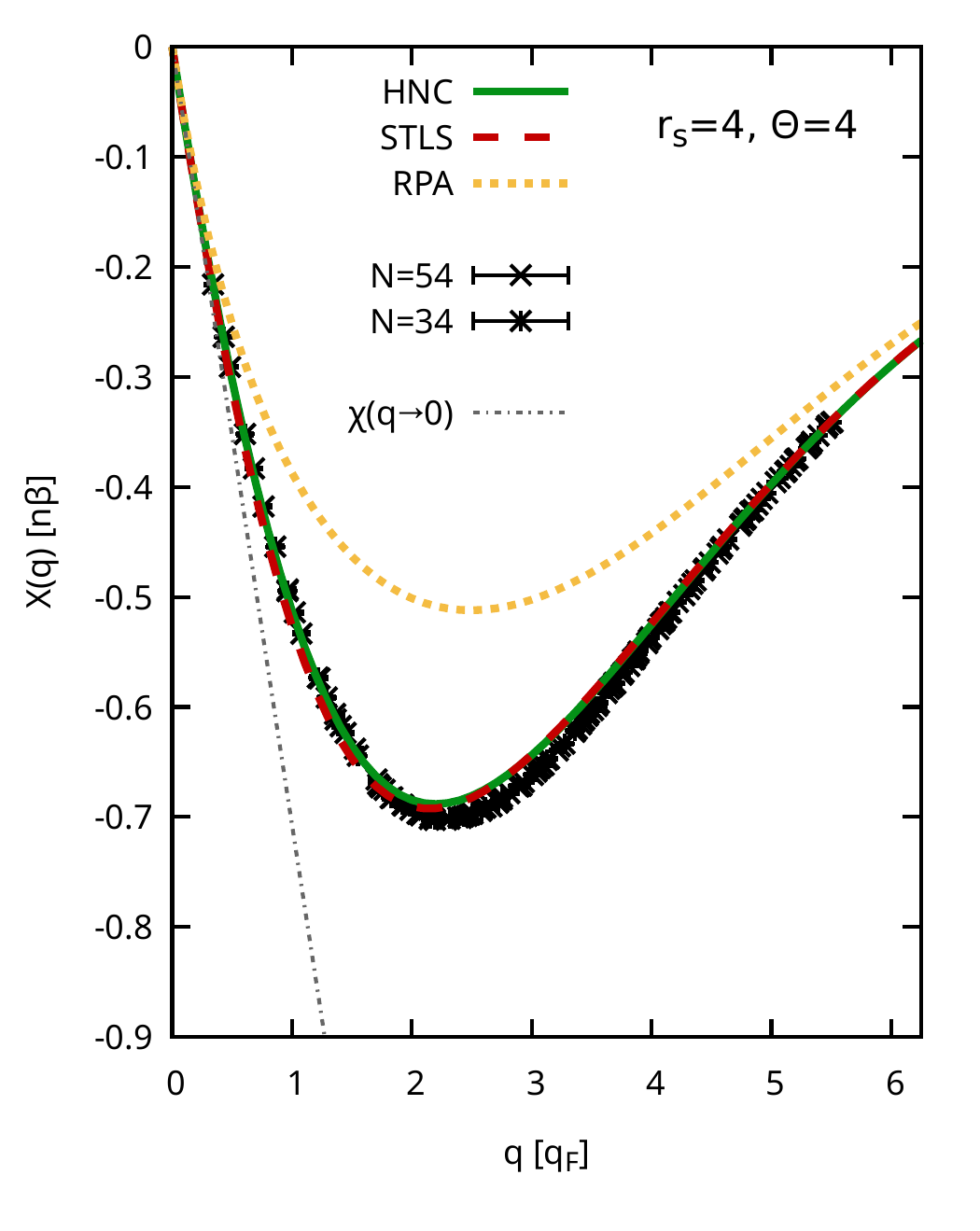}\includegraphics[width=0.315\textwidth]{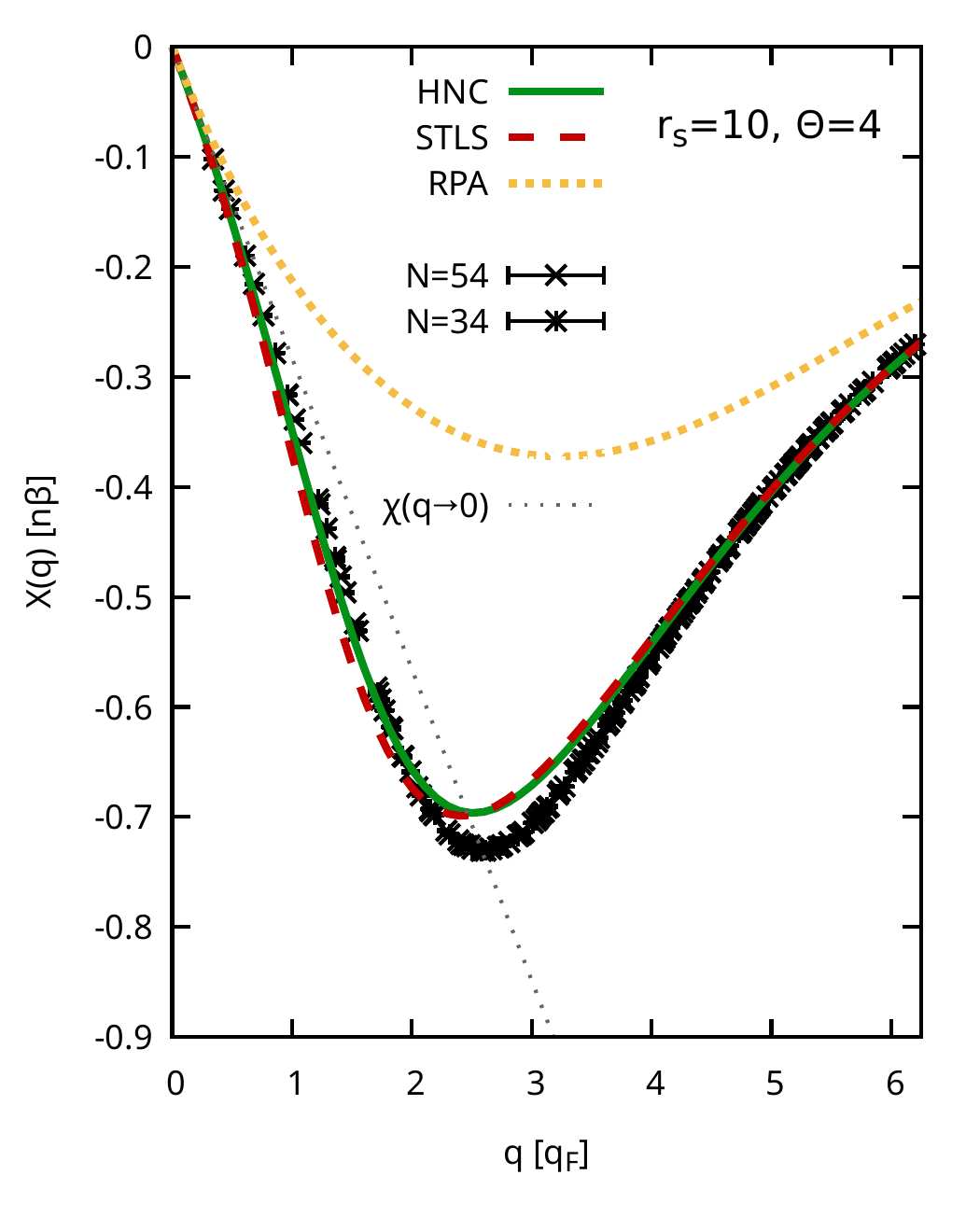}\\\vspace*{-1cm}\includegraphics[width=0.315\textwidth]{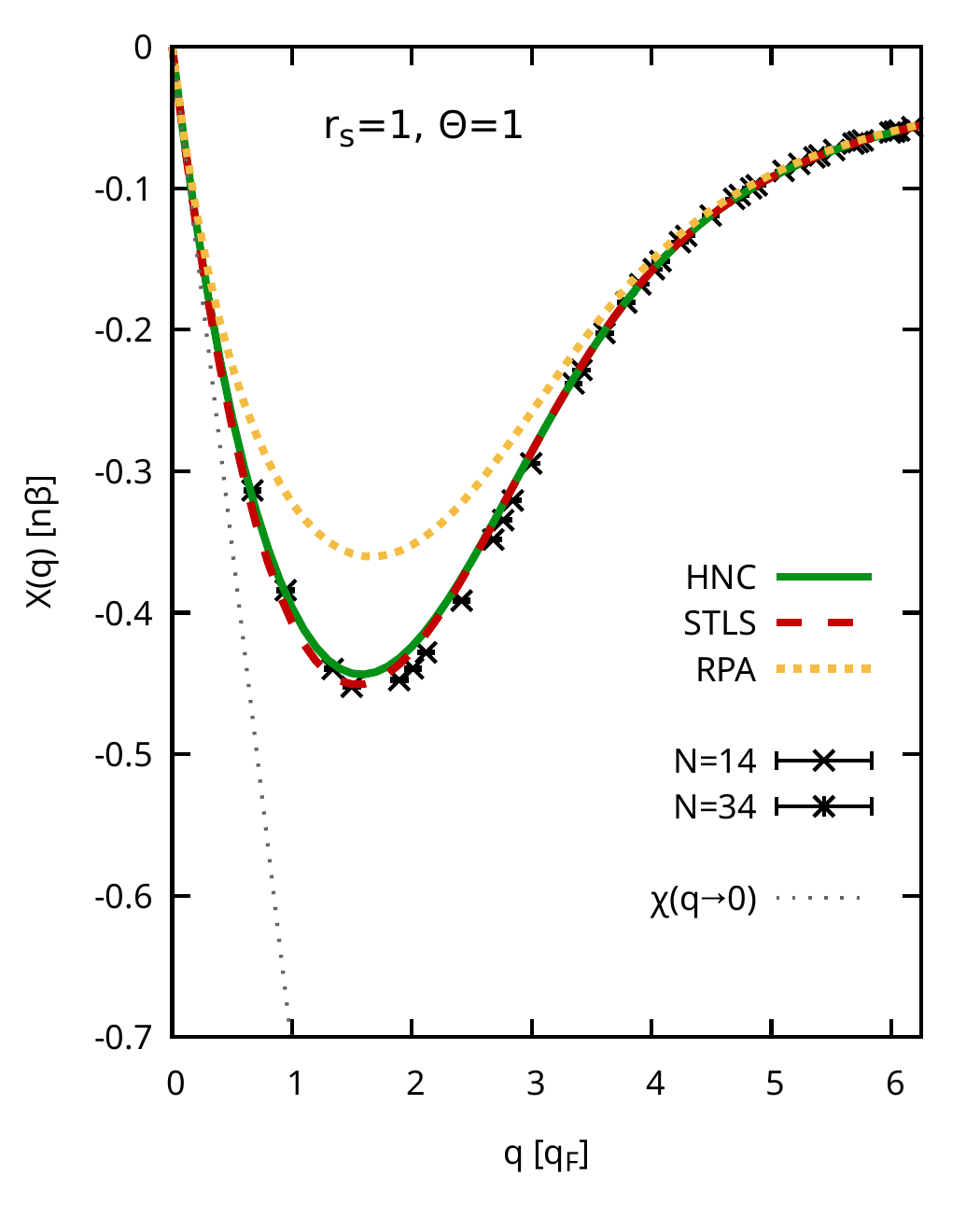}\includegraphics[width=0.315\textwidth]{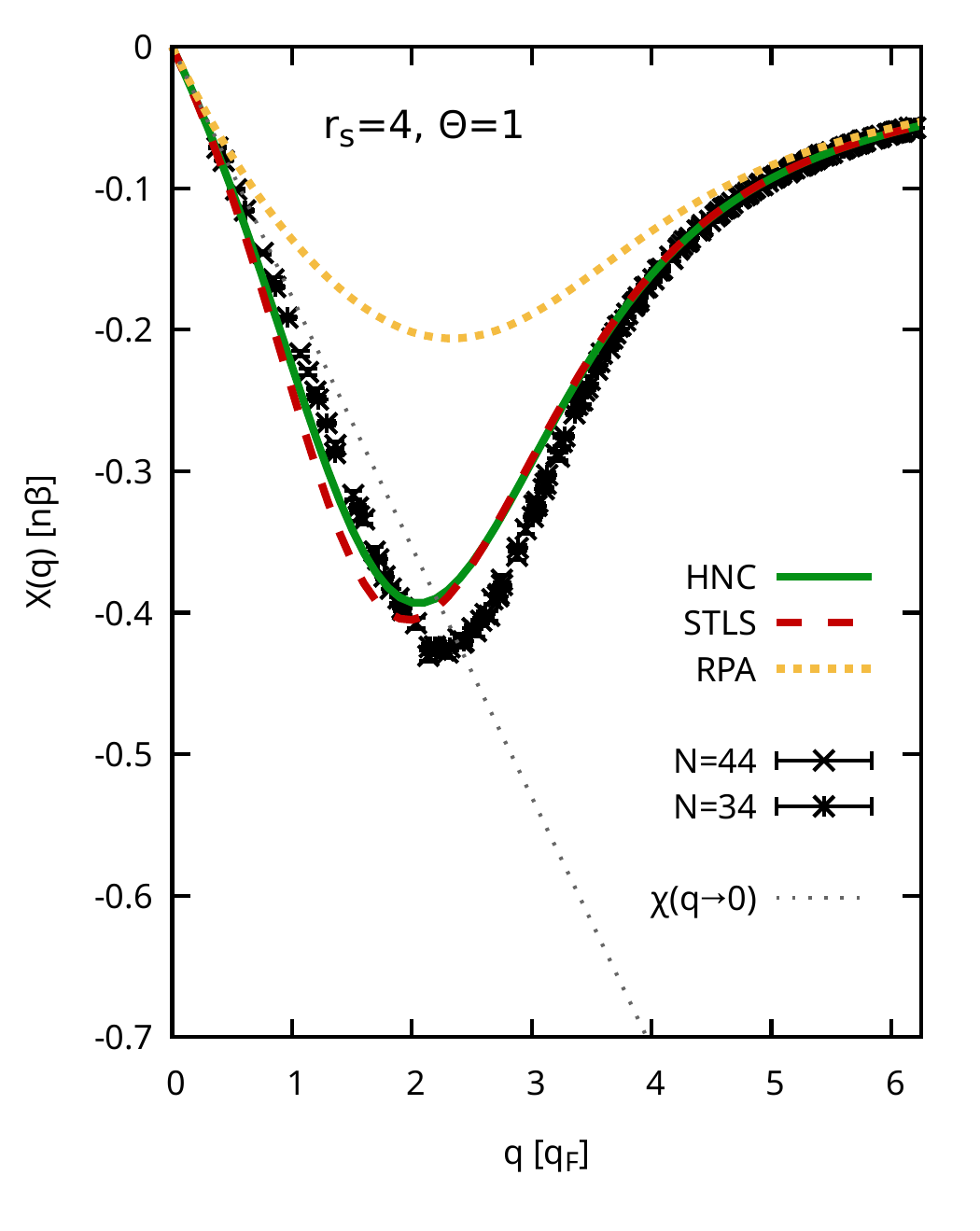}\includegraphics[width=0.315\textwidth]{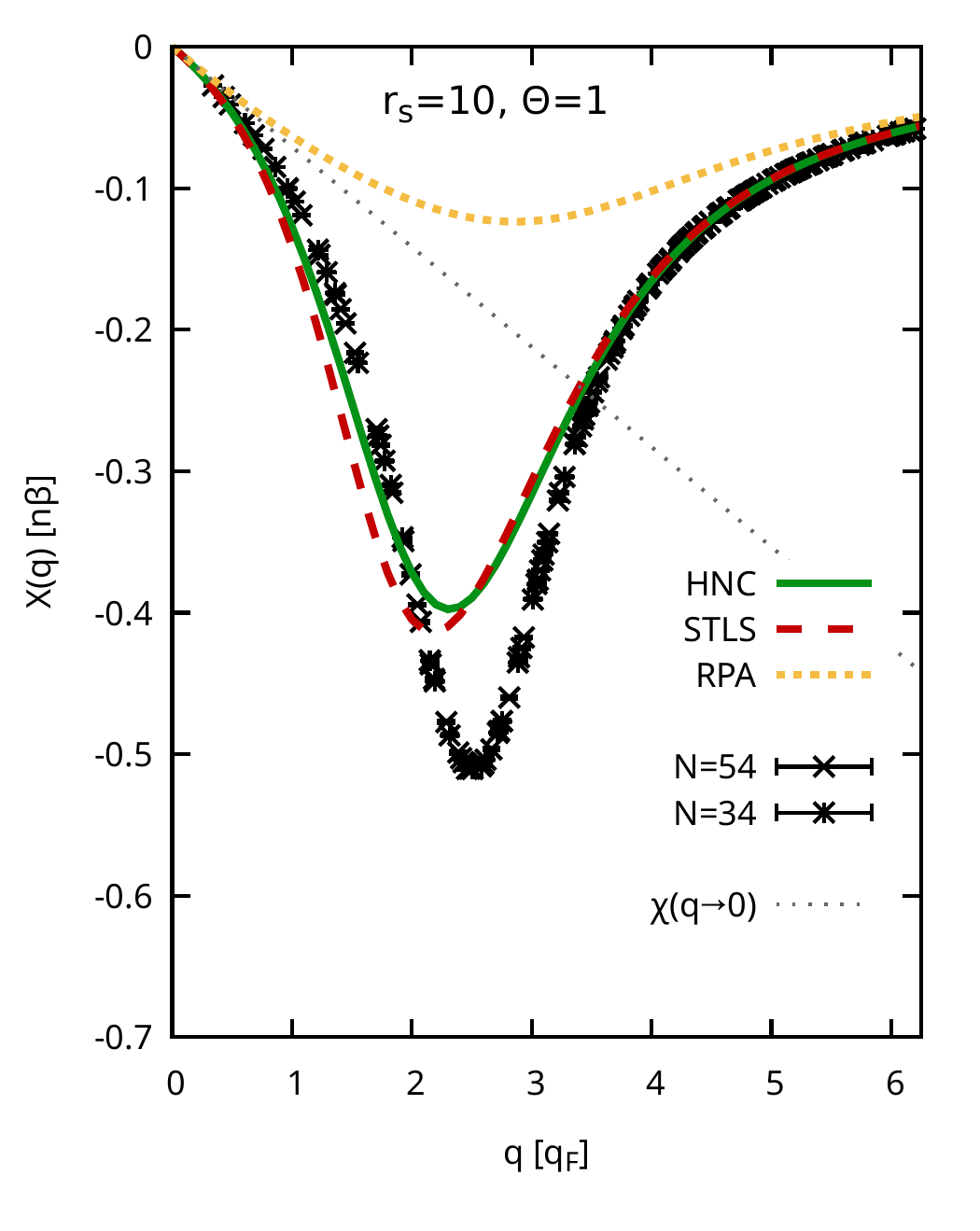}
    \caption{2DEG linear static density response at $r_s=1$ (left), $r_s=4$ (center), and $r_s=10$ (right) with the top and bottom rows corresponding to $\Theta=4$ and $\Theta=1$, respectively. Black symbols: quasi-exact PIMC reference data; solid green: HNC scheme; dashed red: STLS scheme; dotted yellow: RPA. The dotted gray lines correspond to the exact $q\to0$ limit, see Eq.~(\ref{eq:chi0}).
    }
    \label{fig:CHI}
\end{figure*}

\subsection{Static density response\label{sec:chi}}

\noindent The third key observable that we investigate herein concerns the linear static density response function $\chi(\mathbf{q})\equiv\chi(\mathbf{q},\omega=0)\equiv\widetilde{\chi}(\mathbf{q},\ell=0)$.
In Fig.~\ref{fig:CHI}, we compare our new dielectric theory results for $\chi(\mathbf{q})$
in detail to new PIMC data, which we compute from the imaginary-time version of the fluctuation--dissipation theorem~\cite{bowen2,Dornheim_MRE_2023},
\begin{eqnarray}\label{eq:static_chi}
    \chi(\mathbf{q}) = - n \int_0^\beta \textnormal{d}\tau\ F(\mathbf{q},\tau)\ ,
\end{eqnarray}
with $F(\mathbf{q},\tau) = \braket{\hat{n}(\mathbf{q},0)\hat{n}(-\mathbf{q},\tau)}$ being the imaginary-time density--density correlation function and $\tau\in[0,\beta]$ the (reduced) imaginary time; see, e.g., Refs.~\cite{Dornheim_PTR_2023,Dornheim_MRE_2023,Dornheim_MRE_2024,Dornheim_T_2022,Dornheim_moments_2023} for a more detailed discussion of its physical applications. We note that Eq.~(\ref{eq:static_chi}) holds for arbitrary dimensionality and has been used extensively for past investigations of the 3D-UEG~\cite{dynamic_folgepaper,dornheim_ML,dornheim_electron_liquid,Dornheim_PRR_2022} and other WDM systems~\cite{Dornheim_MRE_2024,schwalbe2025staticlineardensityresponse}. Overall, we observe the same level of accuracy of the dielectric results (HNC, STLS and RPA) as for the static structure factor $S(\mathbf{q})$ shown in Fig.~\ref{fig:SSF} above for the same conditions: the RPA is exact for small and large wavenumbers, with
\begin{eqnarray}\label{eq:chi0}
    \chi(q\to0) = - \frac{q}{2\pi}\ ,
\end{eqnarray}
but systematically underestimates the true density response for intermediate $q$; the STLS and HNC schemes constitute significant improvements over the RPA everywhere, but they underestimate the true depth of the minimum in $\chi(\mathbf{q})$ at around $q\approx2.5q_\textnormal{F}$; no depicted dielectric scheme is capable of accurately describing the true density response at $r_s=10$, in particular at $\Theta=1$.

\begin{figure*}
    \centering
    \includegraphics[width=0.485\textwidth]{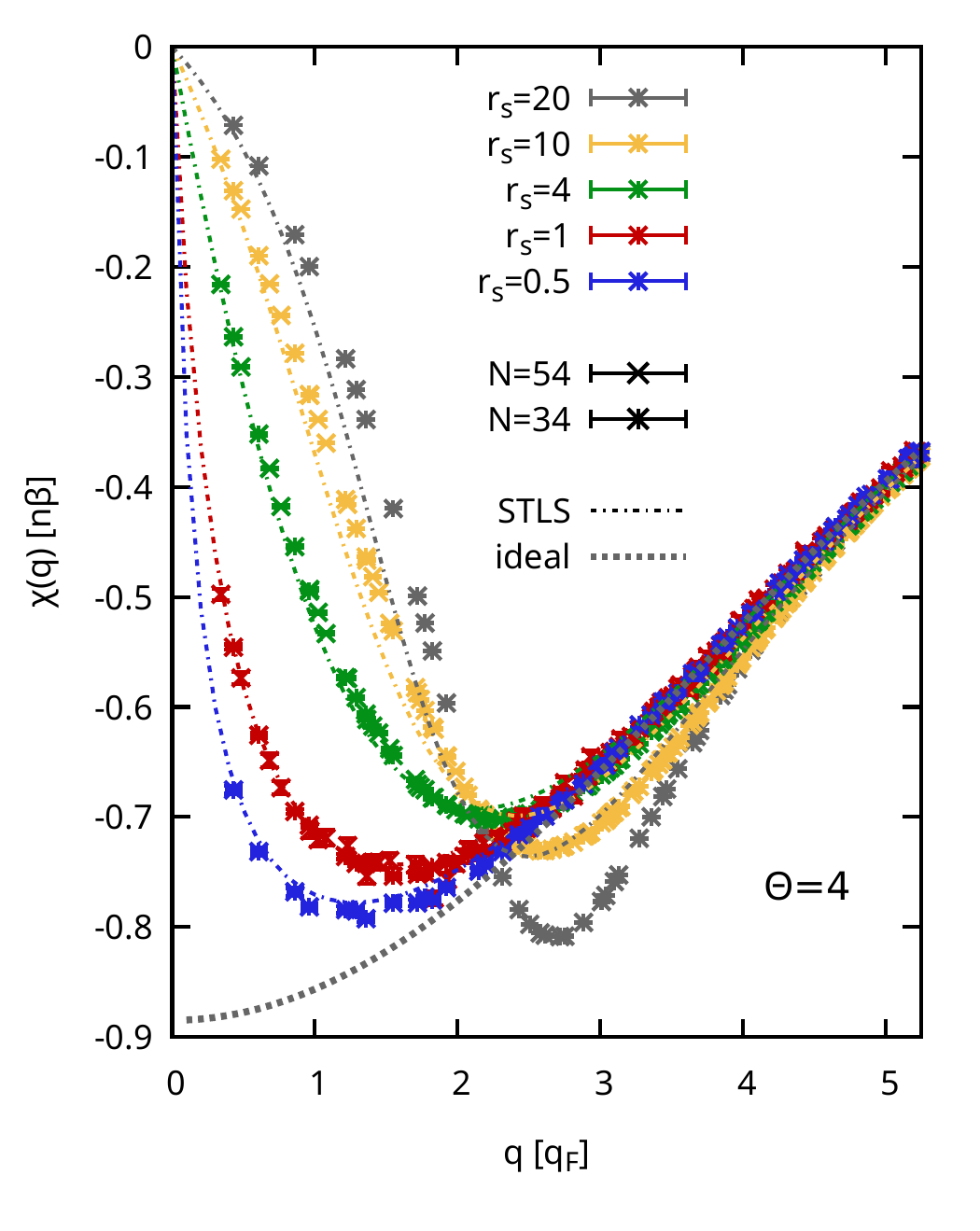}\includegraphics[width=0.485\textwidth]{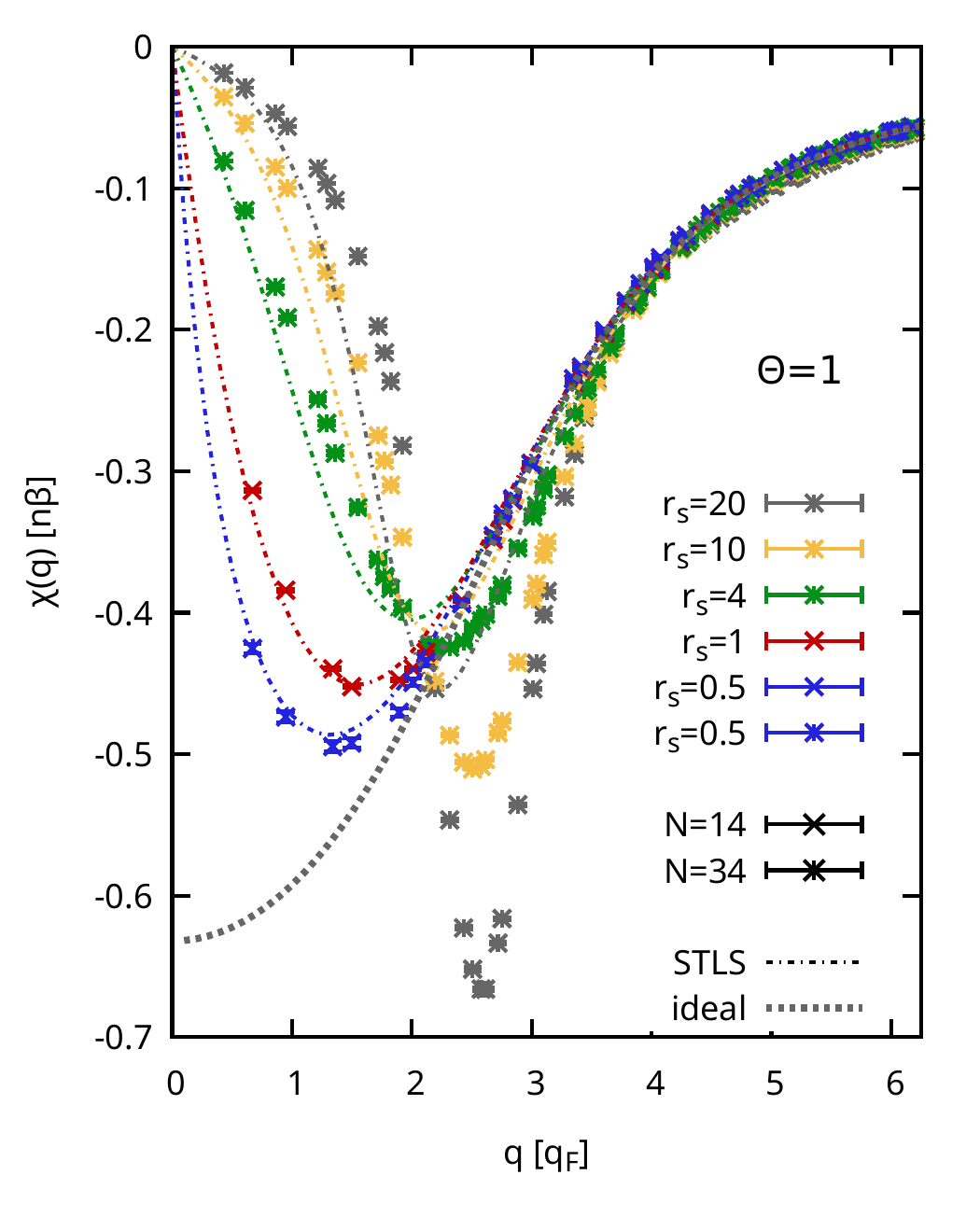}
    \caption{Dependence of the 2DEG linear static density response $\chi(\mathbf{q})$ on the density parameter $r_s$ for $\Theta=4$ (left), $\Theta=1$ (right). The stars and crosses show quasi-exact PIMC reference results for different system sizes $N$. The gray, yellow, green, red, and blue symbols correspond to $r_s=20, 10, 5, 1, 0.5$, respectively. The STLS results are included as the dash-dotted lines with the same color code. The exact ideal Fermi gas results are included as black dashed lines.
    }
    \label{fig:CHI_rs}
\end{figure*}

To investigate the physical meaning and behavior of $\chi(\mathbf{q})$ in more detail, we consider its dependence on the coupling parameter $r_s$ in Fig.~\ref{fig:CHI_rs} for $\Theta=4$ (left) and for $\Theta=1$ (right). The dotted gray lines show the density response of a non-interacting (ideal) Fermi gas at the same conditions, which does not depend on the density for a given value of $\Theta$. First, we note that the ideal response attains a finite value for $q\to0$ owing to the absence of screening effects without Coulomb repulsion. For classical systems, we would simply have $\chi_0(\mathbf{q})\equiv -n\beta$ independent of the wavenumber~\cite{Kugler1975}. For quantum systems, on the other hand, $\chi_0(\mathbf{q})$ monotonically decreases with $q$ due to quantum delocalization effects~\cite{Dornheim_Nature_2022,Dornheim_MRE_2024,Dornheim_review}. Indeed, the electrons will cease to respond to an external perturbation when the thermal de Broglie wavelength $\lambda_T=\sqrt{2\pi\hbar^2\beta/m_e}$ is much larger than the perturbation wavelength $\lambda=2\pi/q$. This holds equally for the ideal Fermi gas and  interacting UEG, which converge towards each other in the single particle limit of $q\gg q_\textnormal{F}$.
In practice, the impact of quantum delocalization can also be gleamed intuitively from Eq.~(\ref{eq:static_chi}): in reduced units, the static density response is simply given by the area under $F(\mathbf{q},\tau)$, which exhibits a rapid decay with $\tau$ for large $q$. This trend can already be seen from the well known f-sum rule, which states that the first derivative of $F(\mathbf{q},\tau)$ around $\tau=0$ increases parabolically with $q$~\cite{Dornheim_moments_2023,Dornheim_SciRep_2024}.

In contrast, the density response of the interacting 2DEG exhibits a significantly more interesting behavior. First, $\chi(\mathbf{q})$ is non-monotonic for all depicted conditions, vanishing in the $q\to0$ and $q\to\infty$ limits with a pronounced negative minimum in between. Interestingly, the depth and position of this minimum [note that we show $\chi(\mathbf{q})$ in units of $n\beta$ to make the curves directly comparable] significantly depend on $r_s$, which is the outcome of a competition between two opposing effects. On the one hand, the stronger coupling at larger $r_s$ energetically rewards the alignment with an external perturbation when the wavelength is commensurate with the average interparticle distance. In that case, the system strongly responds to the perturbation, which directly explains the more pronounced response at $r_s=20$, in particular at $\Theta=1$. We point out that this effect is also directly related to the roton-type excitation in the dynamic structure factor, which has been studied intensely for the 3D-UEG~\cite{Dornheim_Nature_2022,Dornheim_Force_2022,Chuna_JCP_2025,Takada_PRB_2016,koskelo2023shortrange}. On the other hand, the static density response is less strongly screened for small $q$ [note that we consider $q$ in units of $q_\textnormal{F}$ here, making Eq.~(\ref{eq:chi0}) depend on the density], which allows it to more closely follow $\chi_0(\mathbf{q})$. As a consequence, the position of the minimum is shifted to smaller $q$ with decreasing $r_s$, and the reduced screening allows for deeper minima.

Our new STLS results nicely capture the reduced screening at high densities, but fail to reproduce the exchange--correlation induced formation of the deeper minimum at intermediate $q$ for strong coupling.

\subsection{Interaction energy and exchange correlation free energy}\label{sec:uintfxc}

\noindent We next turn to the thermodynamic quantities that can be obtained directly from the SSF $S(\mathbf{q})$, namely the interaction energy and the exchange--correlation (XC) free energy. Specifically, the STLS scheme was solved for the 2DEG at density parameters $r_s\in[0.01,1.0]$ with step size $\Delta r_s=0.001$ and $r_s\in(1.0,10.0]$ with $\Delta r_s=0.01$, combined with reduced temperatures
$\Theta\in[0.01,0.1]$ with $\Delta\Theta=0.01$ and $\Theta\in[0.1,10.0]$ with $\Delta\Theta=0.1$,
amounting to a total of $206\,119$ distinct state points. To the best of our knowledge, this constitutes the most comprehensive STLS dataset for the finite temperature 2DEG available to date. The interaction energy is obtained through $S(\mathbf{q})$ according to~\cite{hansen2013theory}
\begin{equation}
U_{\mathrm{int}} = \frac{1}{2} \int \frac{\mathrm{d}^D k}{(2\pi)^D}\,
U_D(\mathbf{k}) \left[ S(\mathbf{k}) - 1 \right].
\label{eq:uint}
\end{equation}
Employing Hartree energy units and using the same normalizations as in Sec.~\ref{subsub1}, we arrive at the normalized two dimensional interaction energy 
\begin{equation}
     \widetilde{u}_{\mathrm{int}}(r_s,\Theta)=\frac{\sqrt{2}}{2}\frac{1}{r_s}\int_0^{\infty}[S(y;r_s,\Theta)-1]dy .
\end{equation}
From the interaction energy, we subsequently obtain the exchange--correlation free energy $\widetilde{f}_{\mathrm{xc}}(r_s,\Theta)$ from the standard coupling-constant integration~\cite{quantum_theory}
\begin{equation}
    \widetilde{f}_{\mathrm{xc}}(r_s,\Theta)
    = \frac{1}{r_s^2} \int_{0}^{r_s} r_s^{\prime} \widetilde{u}_{\mathrm{int}}(r_s',\Theta)\, \mathrm{d}r_s'.\label{eq:fxc}
\end{equation}
Differentiation of Eq.~(\ref{eq:fxc}) with respect to $r_s$ leads to the differential form of the adiabatic connection formula
\begin{equation}
    \widetilde{u}_{\mathrm{int}}(r_s,\Theta)
    = 2 \widetilde{f}_{\mathrm{xc}}(r_s,\Theta)
    + r_s \frac{\partial \widetilde{f}_{\mathrm{xc}}(r_s,\Theta)}{\partial r_s} .
    \label{eq:placeholder_label}
\end{equation}
Owing to the dense and highly resolved $(r_s,\Theta)$ grid comprising more than $2\times10^5$ 2DEG state points, the resulting exchange--correlation free energy data are sufficiently smooth and accurate to permit a reliable global parametrization. We therefore choose $\widetilde{f}_{\mathrm{xc}}$ as the primary fitting target, as it is the thermodynamically fundamental quantity entering, e.g., finite--temperature DFT. 

The high--density limit of the xc free energy is governed by the Hartree--Fock contribution, which can be numerically calculated with arbitrary precision and which we explicitly enforce in the fit via the prefactor $\alpha_{\mathrm{HF}}(\Theta)$~\cite{Toliasmisc}. For this quantity, we employ the Pad{\'e} parametrization
\begin{equation*}
    \begin{split}
    \alpha_{\text{HF}}(\Theta)=&\frac{\sqrt{2}}{\pi}\tanh\Big(\frac{1}{\sqrt{\Theta}}\Big)\times\\&\frac{\frac{4}{3}+a_{1/2}\Theta^{1/2}+a_1\Theta+a_{3/2}\Theta^{3/2}+a_{2}\Theta^{2}}{1+b_{1/2}\Theta^{1/2}+b_{1}\Theta^{}+b_{3/2}\Theta^{3/2}+b_{2}\Theta^{2}}.
\end{split}
\end{equation*}
with the coefficients being shown in Table~\ref{Table:aHF}.

\begin{table}
    \caption{Pad{\'e} coefficients for the parametrization of the 2DEG Hartree--Fock pre-factor $\alpha_{\mathrm{HF}}(\Theta)$.}
    \centering
    \begin{tabular}{cc}
    \hline
     $a$--coefficients & $b$--coefficients \\
    \hline
     $a_{1/2} = -2.04423683712$ & $b_{1/2} = -1.5526945699$ \\
     $a_{1}   =  5.71731878382$ & $b_{1}   =  4.57339734818$ \\
     $a_{3/2} = -2.94589021064$ & $b_{3/2} = -2.98706116473$ \\
     $a_{2}   =  6.69196206698$ & $b_{2}   =  6.79835205067$ \\
    \hline
    \end{tabular}
    \label{Table:aHF}
\end{table}

\begin{figure*}
    \centering
    \includegraphics[width=0.485\textwidth]{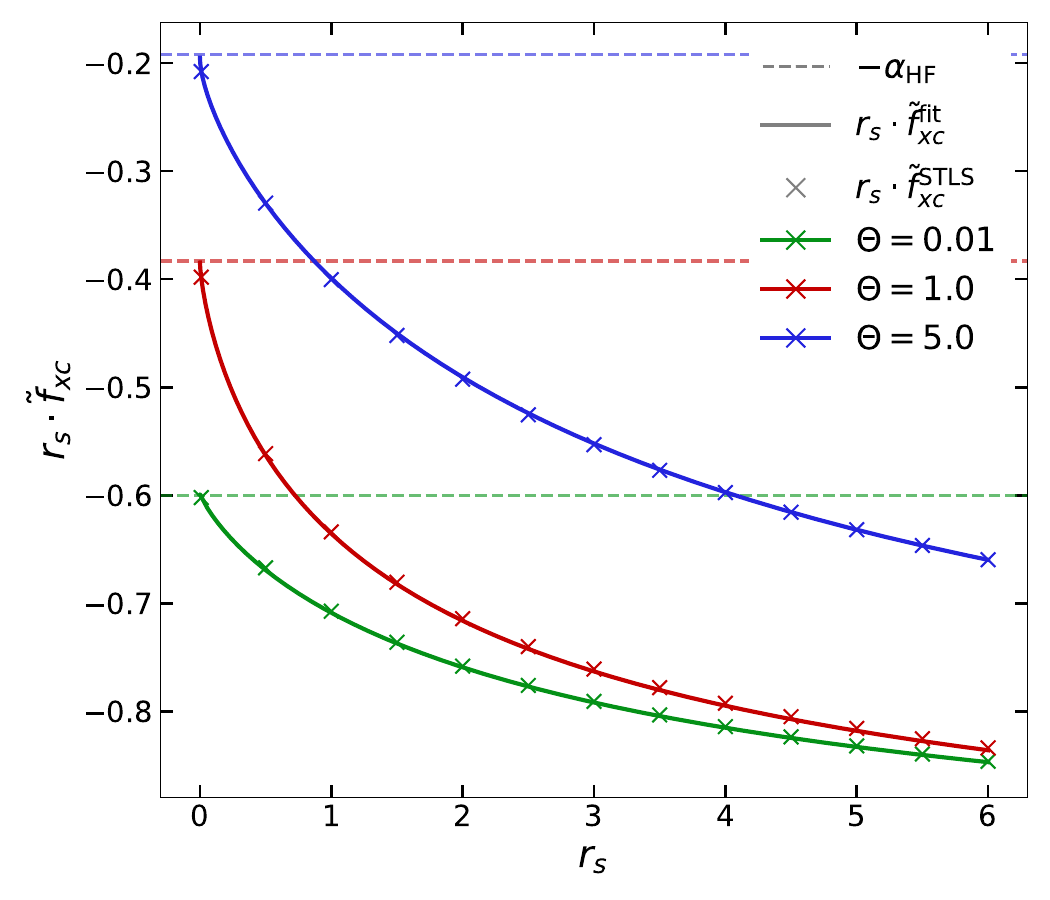}\includegraphics[width=0.485\textwidth]{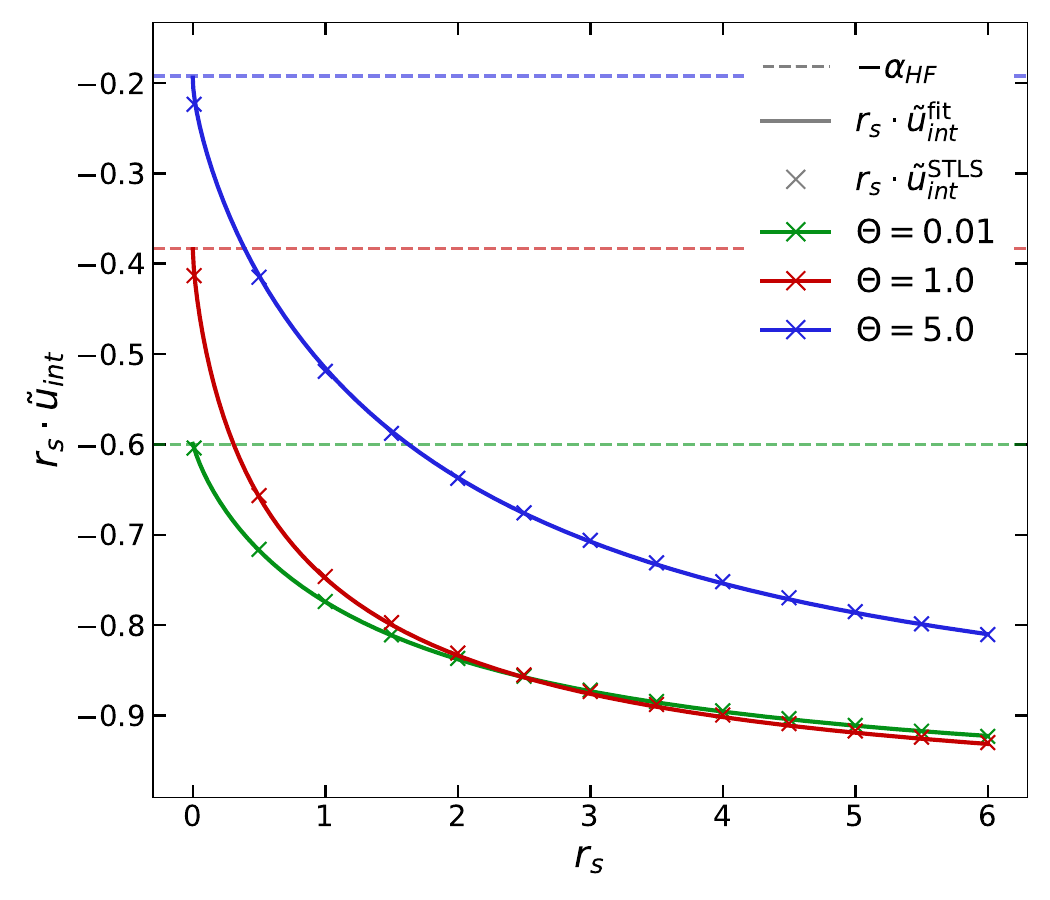}
    \caption{Comparison between the raw 2DEG STLS thermodynamic data (crosses) and their analytical parametrizations (solid lines) as functions of $r_s$ for selected degeneracy parameters $\Theta$. The high--density Hartree--Fock limits (dashed horizontal lines) are also presented. Left panel: rescaled exchange correlation free energy $r_s\,\widetilde{f}_{\mathrm{xc}}(r_s,\Theta)$. Right panel: rescaled interaction energy $r_s\,\widetilde{u}_{\mathrm{int}}(r_s,\Theta)$ obtained directly from the static structure factor integration and via differentiation of the fitted $\widetilde{f}_{\mathrm{xc}}$.
}    \label{fig:rsfxcuint}
\end{figure*}

To represent the full $(r_s,\Theta)$ dependence of the xc free energy, we adopt the rational form given in Eq.~(\ref{fxcfitgen}). The temperature dependence of the fit coefficients is encoded through Pad{\'e} forms of the type shown in Eq.~(\ref{fxcfitpade}), ensuring regular behavior in both the low--degeneracy and the high--degeneracy limits. All fit parameters are determined from a global nonlinear least--squares fit over the full dataset. The resulting coefficients are summarized in Table~\ref{Table:fxc}.
\begin{equation}
\widetilde{f}^{\text{fit}}_{\mathrm{{xc}}}(r_{\mathrm{s}},\Theta)=-\frac{1}{r_{\mathrm{s}}}\frac{\alpha_{\mathrm{HF}}(\Theta)+\gamma^{\mathrm{f}}(\Theta)r_{\mathrm{s}}^{1/4}+\delta^{\mathrm{f}}(\Theta)r_{\mathrm{s}}^{3/4}}{1+\zeta^{\mathrm{f}}(\Theta)r_{\mathrm{s}}^{1/4}+\eta^{\mathrm{f}}(\Theta)r_{\mathrm{s}}^{3/4}}\,,\label{fxcfitgen}
\end{equation}
\begin{equation}
\kappa^{\mathrm{f}}(\Theta)=\frac{\kappa^{\mathrm{f}}_1+\kappa^{\mathrm{f}}_2\Theta^2+\kappa^{\mathrm{f}}_3\Theta^4+\kappa^{\mathrm{f}}_4\Theta^6}{1+\kappa^{\mathrm{f}}_5\Theta^2+\kappa^{\mathrm{f}}_6\Theta^4+\kappa^{\mathrm{f}}_7\Theta^6}\,.\label{fxcfitpade}
\end{equation}

\begin{table}
    \caption{Pad{\'e} coefficients for the parametrization of the STLS generated 2DEG exchange correlation free energy.}
    \centering
    \begin{tabular}{ccccc}
    \hline
    $i$ & $\gamma_i$ & $\delta_i$ & $\zeta_i$ & $\eta_i$ \\ \hline
    $1$ & \,\,\,-0.138388\,\,\, & \,\,\, 0.337985\,\,\, & \,\,\,-0.216089\,\,\, & \,\,\, 0.344609\,\,\,  \\
    $2$ & \,\,\, 4.366673\,\,\, & \,\,\, 8.792716\,\,\, & \,\,\, 7.111827\,\,\, & \,\,\, 8.071073\,\,\,  \\
    $3$ & \,\,\,-3.957826\,\,\, & \,\,\, 6.857318\,\,\, & \,\,\,-7.693699\,\,\, & \,\,\, 7.048795\,\,\,  \\
    $4$ & \,\,\,-0.013038\,\,\, & \,\,\, 0.049065\,\,\, & \,\,\,-1.001149\,\,\, & \,\,\, 0.100188\,\,\,  \\
    $5$ & \,\,\, 8.979623\,\,\, & \,\,\, 8.881116\,\,\, & \,\,\, 11.539687\,\,\, & \,\,\, 8.178041\,\,\,  \\
    $6$ & \,\,\, 127.614046\,\,\, & \,\,\, 14.400170\,\,\, & \,\,\, 63.611046\,\,\, & \,\,\, 14.529262\,\,\,  \\
    $7$ & \,\,\, 0.956863\,\,\, & \,\,\, 0.926265\,\,\, & \,\,\, 2.111593\,\,\, & \,\,\, 0.992106\,\,\,  \\
    \hline
    \end{tabular}
     \label{Table:fxc}
\end{table}

The resulting parametrization is presented on the left panel of Fig.\ref{fig:rsfxcuint} where we have under-sampled the obtained STLS data in order to better illustrate the parametrization curve. As an internal consistency check, shown on the right of Fig.\ref{fig:rsfxcuint}, we computed the interaction energy $\widetilde{u}_{\mathrm{int}}$ from the fitted exchange--correlation free energy by using Eq.~(\ref{eq:placeholder_label}), and compared with the originally obtained interaction energy data. The overall relative accuracy of the parametrization is $0.08\%$, with no individual state point exceeding a deviation of $0.12\%$. This level of agreement is uniform across the full $(r_s,\Theta)$ domain and confirms that the chosen functional form is sufficiently flexible to represent the STLS thermodynamic data without loss of fidelity.

Finally, we comment on the nontrivial temperature dependence of the interaction energy observed on the right panel of Fig.\ref{fig:rsfxcuint} where curves corresponding to $\Theta=0.01$ and $1.0$ intersect at intermediate coupling. The interaction energy reflects the combined influence of exchange, correlation, and thermal motion, whose relative importance varies strongly with both density and temperature. At weak coupling, exchange dominates and a finite reduced temperature indeed leads to a more negative interaction energy. In contrast, at larger coupling, Coulomb-induced spatial correlations become increasingly important and can outweigh exchange effects. In this strongly coupled regime, quantum effects enter mainly via single-particle delocalization that is associated with the thermal de Broglie wavelength. At very low temperatures, the large de Broglie wavelength leads to an effective delocalization of particles, which counteracts Coulomb-driven localization and reduces the magnitude of the interaction energy. With increasing temperature, the de Broglie wavelength decreases and particles become more localized in this sense, allowing them to sample more strongly correlated configurations. The resulting competition between quantum delocalization and correlation-driven localization provides a plausible explanation for the observed curve crossing. Similar non-monotonic temperature trends and crossover phenomena have been reported in finite temperature 3D-UEG studies~\cite{Dornheim_PRB_2016,Karasiev_2014}.

\section{Summary and Outlook\label{sec:outlook}}

In this work, we have investigated the finite temperature two-dimensional uniform electron gas using two prototypical schemes of the self-consistent dielectric formalism and we have benchmarked the theoretical results against new quasi-exact path integral Monte Carlo data. In particular, we have analyzed the static structure factor, pair correlation function, static density response function, interaction energy, and exchange--correlation free energy across a broad range of densities and temperatures, thereby providing a systematic assessment of dielectric theories in reduced dimensionality. In addition, we have introduced accurate parametrizations for the interaction energy and exchange--correlation free energy, which are internally consistent and directly applicable to finite temperature density functional theory. 

Overall, the present work establishes a comprehensive finite temperature reference for the 2DEG and provides a solid foundation for future investigations of static and dynamic correlation effects in low-dimensional quantum systems. Our results highlight the versatility of the dielectric formalism as a framework for describing correlated quantum systems. The same conceptual approach can be applied across different dimensionalities and pair interaction potentials, as demonstrated for two dimensional Coulomb interacting systems. This flexibility suggests promising extensions to systems interacting with screened Coulomb or short range potentials, such as dipolar systems or quantum fluids, including  helium~\cite{Aziz_JCP_1991}. 

Several directions for future work naturally emerge from this finite-temperature two-dimensional study that are indicative of the flexibility of the dielectric formalism. One possibility concerns the extension of the present approach to fully quantum dielectric schemes, such as the quantum STLS (qSTLS)~\cite{qstls_original,Tolias_2025} and the quantum hypernetted chain (qHNC)~\cite{Tolias_JCP_2023}, which feature dynamic local field corrections and thus include beyond RPA quantum effects. A second possibility concerns the extension of the present approach to dielectric schemes that impose partial self-consistency, such as the Vashishta-Singwi (VS)~\cite{vs_original,Tolias_PRB_2024} and the quantum VS (qVS)~\cite{Hayashi_1980}, whose local field correction (its static limit) automatically satisfies the compressibility sum rule. Another possibility concerns the improvement of the HNC scheme with the inclusion of a parameterized classical bridge function; a technique that has been demonstrated to largely improve the accuracy of the HNC and the qHNC schemes~\cite{Tolias_JCP_2021,LuccoCastello_2022,Tolias_JCP_2023}. It needs to be pointed out though that the extraction and parametrization of Coulomb bridge functions constitutes a complicated and cumbersome feat that has only been performed in 3D thus far~\cite{YOCP_bridge_2021,OCP_bridge_2022}. Alternatively, scaled-HNC approaches also have the potential to improve the HNC and qHNC schemes~\cite{LuccoCastello_2DEG_2021}, while being less complex.

Another important avenue concerns dynamical properties, since the dielectric formalism naturally provides access to dynamic responses in Matsubara space. In particular, the imaginary-time correlation function and its exact Fourier--Matsubara series representation~\cite{Tolias_JCP_2024} offer a direct route to studying the dynamic response at finite temperature. Recent developments in this direction for three-dimensional systems~\cite{Tolias_JCP_2024,Dornheim_PRB_2024} suggest that analogous 2DEG investigations are both timely and feasible. Finally, it would be highly interesting to explore the possible emergence of roton-like features in the 2DEG dynamic structure factor. The roton excitations are known to occur in the 3D-UEG and in strongly correlated two-dimensional Fermi systems such as helium films~\cite{Godfrin2012}.

Finally, an important avenue concerns the complete parametrization of the PIMC-based exchange-correlation free energy of the paramagnetic 2DEG with respect to the density and temperature, which would constitute the two-dimensional analogue of the GDSMFB equation of state~\cite{groth_prl}. This is currently out of reach, because of (i) the absence of accurate 2DEG PIMC results in the entire phase diagram region of interest due to the fermion sign problem~\cite{review}, (ii) the absence of a robust finite-size correction technique for the 2DEG interaction energy that allows extrapolation to the thermodynamic limit~\cite{dornheim_prl}, (iii) the absence of enough 2DEG PIMC results to perform reliable thermodynamic integration of the interaction energy~\cite{review}, (iv) the absence of direct free energy evaluations for the 2DEG within an extended ensemble PIMC approach~\cite{dornheim2024directfreeenergycalculation,Dornheim_PRR_2025,Svensson_2025,dornheim2025fermionicfreeenergiestextitab}. Apart from its apparent utility, such a quasi-exact equation of state would also allow us to quantify the accuracy and systematic errors of different dielectric schemes and quantum-classical mapping approaches.

\begin{figure*}
    \centering
    \includegraphics[width=0.391\textwidth]{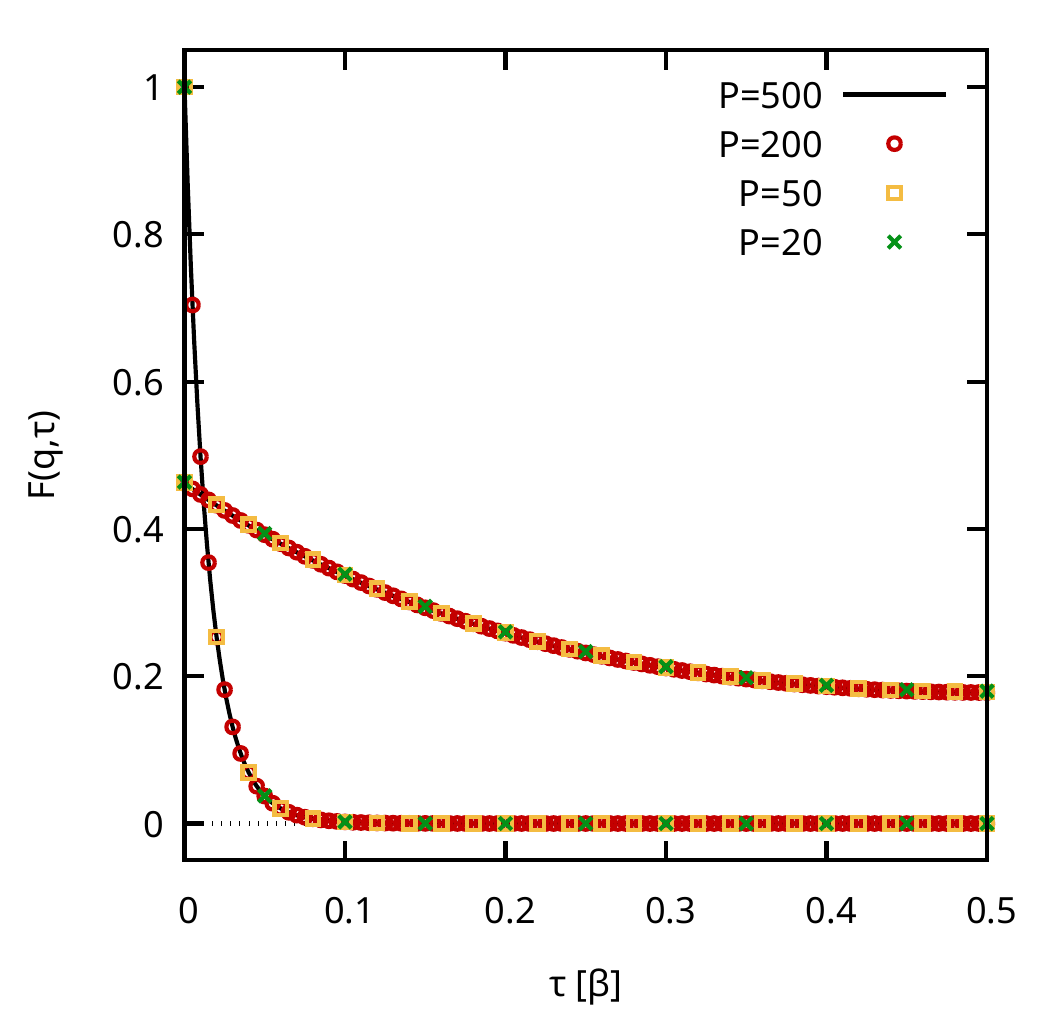}\includegraphics[width=0.391\textwidth]{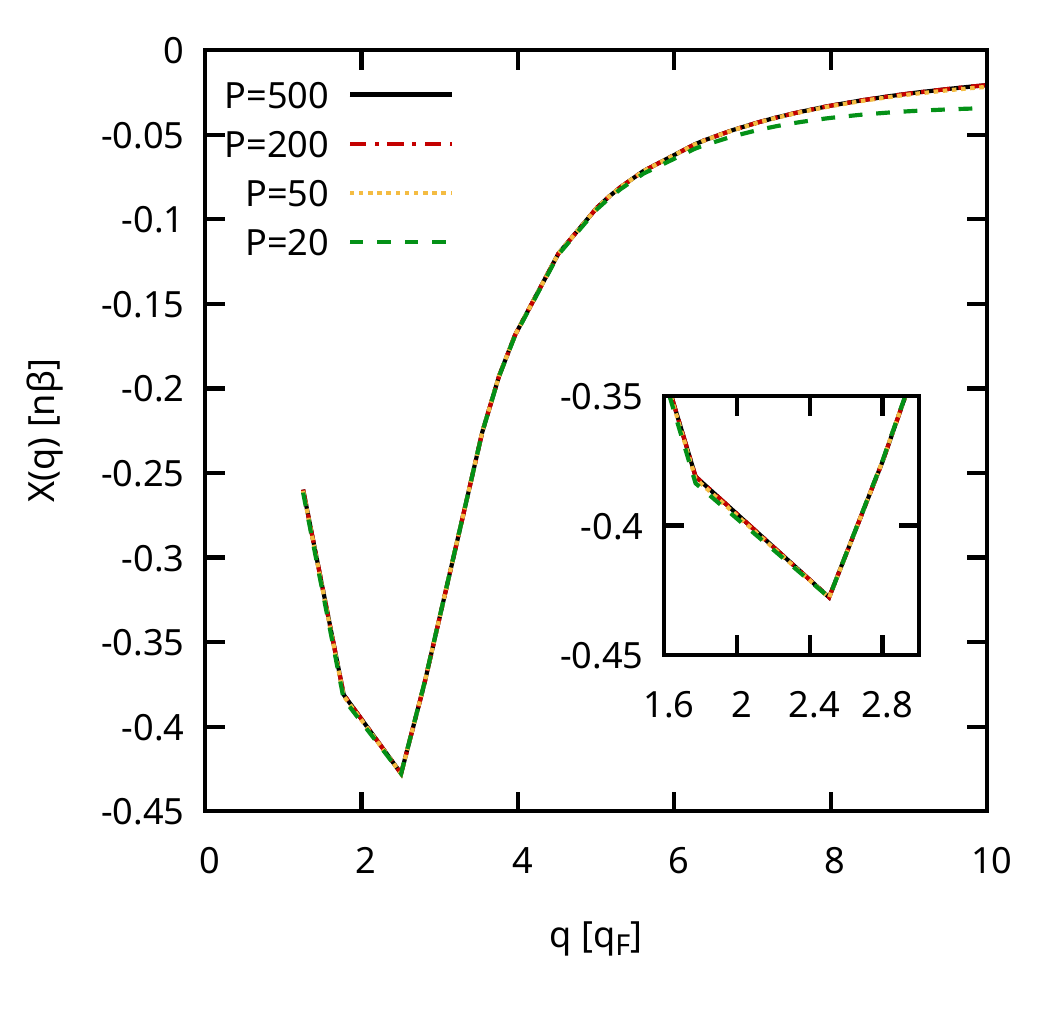}
    \caption{Convergence of the \emph{ab initio} PIMC results with the number of imaginary time propagators $P$ for the imaginary-time correlation function $F(\mathbf{q},\tau)$ (left) and static density response $\chi(\mathbf{q})$ computed via Eq.~(\ref{eq:static_chi}) (right) for $N=4$ at $r_s=4$ and $\Theta=1$.
    }
    \label{fig:PIMC_propagators}
\end{figure*}

\appendix
\section{Computation of the STLS and HNC LFC for the 2DEG\label{appendix:STLS_HNC}}

\noindent We start from the \emph{general STLS LFC functional} for arbitrary pair interaction and dimensionality,
\begin{equation*}
G_D^{\mathrm{STLS}}(\mathbf{q})
=
-\frac{1}{n}
\int \frac{d^Dk}{(2\pi)^D}
\frac{U_D(k)}{U_D(q)}
\frac{\mathbf{q}\cdot\mathbf{k}}{q^2}
\left[ S(\mathbf{q}-\mathbf{k}) - 1 \right].
\end{equation*}
After introducing the two-dimensional Coulomb interaction $U_{2D}(q)=2\pi e^2/q$, defining the normalized variables $\mathbf{x}=\mathbf{q}/q_F$ and $\mathbf{y}=\mathbf{k}/q_F$, orienting $\mathbf{q}$ along the $x$ axis without loss of generality, we switch to polar coordinates and exploit the isotropy of $S(\mathbf{q})$, which yields
\begin{equation}
\begin{split}
    G_{2D}^{\mathrm{STLS}}(x)
=
-\frac{1}{2\pi}
\int_0^\infty dy\, y [S(y)-1]\times\\
\int_0^{2\pi} d\phi\,
\frac{x - y\cos\phi}{\sqrt{x^2 + y^2 - 2xy\cos\phi}} .
\label{eq:stls_ang}
\end{split}
\end{equation}
The angular kernel in Eq.~(\ref{eq:stls_ang}) cannot be reduced to elementary functions.
By exploiting the symmetry $\phi\to2\pi-\phi$ and applying the trigonometric substitution $\phi=2\theta$, the integral can be expressed in terms of complete elliptic integrals.
After some cumbersome algebra, one finds
\begin{align*}
\begin{split}
\int_0^{2\pi} d\phi\,
\frac{x - y\cos\phi}{\sqrt{x^2 + y^2 - 2xy\cos\phi}}
=
\frac{2}{x}
\Big[
(x-y)\mathcal{K}(m) + \\(x+y)\mathcal{E}(m)
\Big],
\end{split}
\end{align*}
where the modulus is given by $m=2\sqrt{xy}/(x+y)$. Insertion of this identity into Eq.~(\ref{eq:stls_ang}) leads directly to the representation given in Eq.~(\ref{eq:STLS_LFC}).

We start from the \emph{general HNC LFC functional} for arbitrary pair interaction and dimensionality, that features an extra nonlinear coupling between $G(\mathbf{q})$ and $S(\mathbf{q})$,
\begin{align*}
    G_D^{\text{HNC}}(\mathbf{q}) =& -\frac{1}{n} \int \frac{d^Dk}{(2\pi)^D}  \frac{U_D(\mathbf{k})}{U_D(\mathbf{q})}\frac{\mathbf{q} \cdot \mathbf{k}}{q^2}\times\\
    &[S(\mathbf{q}-\mathbf{k}) - 1] \{1 - [G(\mathbf{k}) - 1][S(\mathbf{k}) - 1]\}.
\end{align*}
After the same normalization as above and the introduction of polar coordinates, the problem reduces to a two-dimensional integral over $y$ and $\phi$ with a kernel containing square-root singularities similar to the STLS case. To obtain a tractable form, we introduce the change of variables
\begin{equation*}
u = \sqrt{y^2 + \frac{x^2}{4} + yx\cos\phi},
\qquad
w = \sqrt{y^2 + \frac{x^2}{4} - yx\cos\phi},
\end{equation*}
which map the integration domain onto a triangular region defined by
$|u-w|\le x \le u+w$. The Jacobian determinant of this transformation,
\begin{equation*}
\left|\frac{\partial(y,\phi)}{\partial(u,w)}\right|
=
\frac{wu}{x\sqrt{4u^2x^2-(u^2-w^2+x^2)^2}},
\end{equation*}
remains finite over the reduced polar interval. The HNC LFC can easily be decomposed to the STLS LFC plus an LFC corrective term, which only becomes important at moderate-to-strong coupling. This is the form of Eq.~(\ref{eq:HNC_LFC}).

\section{Convergence of the PIMC results\label{appendix:PIMC}}

\noindent In order to confirm the quasi-exactness of our \emph{ab initio} PIMC results, we must ensure that the factorization error of the thermal density matrix vanishes within the given Monte Carlo error bars. In the left panel of the Fig.~\ref{fig:PIMC_propagators}, we plot the imaginary-time density--density correlation function $F(\mathbf{q},\tau)$ of the 2DEG versus the imaginary time at $r_s=4$, $\Theta=1$ for two selected values of the wavenumber $q$ with $N=4$. The solid black line has been obtained for $P=500$ and it serves as a baseline; the red circles, yellow squares and green crosses correspond to $P=200$, $P=50$, and $P=20$, respectively. Clearly, no dependence on $P$ can be resolved with the bare eye on the depicted scale.

In the right panel of Fig.~\ref{fig:PIMC_propagators}, we show the $q$-dependence of the static density response at the same conditions. Here, we find very small deviations from the other curves for $P=20$ at moderate $q$ (see the inset showing a magnified segment) and larger deviations at large $q$.
The latter is a well-known discretization error when evaluating Eq.~(\ref{eq:static_chi}) due to the finite $\tau$-grid, and not predominantly a factorization error. We safely conclude that $P=200-500$, which have been used for all results of the main text, are sufficient to ensure convergence with respect to $P$.

\begin{acknowledgements}

\noindent This work was supported by the Center for Advanced Systems Understanding (CASUS), financed by Germany’s Federal Ministry of Education \& Research and the Saxon state government out of the budget approved by the Saxon State Parliament. Support is also acknowledged for the CASUS Open Project \emph{Guiding dielectric theories with ab initio quantum Monte Carlo simulations: from the strongly coupled electron liquid to warm dense matter}. This work also received funding from the European Research Council (ERC) under European Union’s Horizon 2022 research and innovation programme (Grant agreement No. 101076233, "PREXTREME"). The views and opinions expressed are however those of the authors only and do not necessarily reflect those of the European Union or the European Research Council Executive Agency. Neither the European Union nor the granting authority can be held responsible for them. Tobias Dornheim gratefully acknowledges funding from the Deutsche Forschungsgemeinschaft (DFG) via project DO 2670/1-1. The computations were performed on a Bull Cluster at the Center for Information Services and High-Performance Computing (ZIH) at Technische Universit\"at Dresden and at the Norddeutscher Verbund f\"ur Hoch- und H\"ochstleistungsrechnen (HLRN) under grant mvp00024.

\end{acknowledgements}

\bibliography{bibliography}

\end{document}